\documentstyle[12pt,amsfonts]{article}

\def\half{\textstyle{\frac{1}{2}}}

\def\H{{\cal H}}
\def\p{\phi}
\def\B{\beta}
\def\g{\gamma}
\def\kk{k^{\sf plus}}
\def\l{\lambda}

\def\S{\Sigma'}

\def\ra{\rightarrow}
\def\tint{{\textstyle\int}}

\def\hp{{\hat\pi}}
\def\hph{{\hat\phi}}
\def\s{\hskip.08em}
\def\P{\Pi'}
\def\d{\partial}

\def\a{\alpha}
\def\b{\begin{eqnarray*}}  
\def\e{\end{eqnarray*}}    
\def\bn{\begin{eqnarray}}  

\def\<{\langle}
\def\>{\rangle}

\def\k{\kappa}
\def\{{\lbrace}
\def\}{\rbrace}
\begin{document}
\title{A New Approach to \\Nonrenormalizable  Models}
\author{John R.~Klauder\\
Department of Physics and Department of Mathematics\\
University of Florida\\
Gainesville, FL 32611 ~~USA\footnote{Permanent address}\\and\\
Department of Physics\\
The Norwegian University of Science and Technology\\
 N-7491 Trondheim,
Norway }

\date{}    
\maketitle
\begin{abstract}
Nonrenormalizable quantum field theories require counterterms; and
based on the hard-core interpretation of such interactions, it is
initially argued, contrary to the standard view, that counterterms
suggested by renormalized perturbation theory are in fact
inappropriate for this purpose. Guided by the potential underlying
causes of triviality of such models, as obtained by alternative
analyses, we focus attention on the ground-state distribution
function, and suggest a formulation of such distributions that
exhibits nontriviality from the start. Primary discussion is focused
on self-interacting scalar fields. Conditions for bounds on general
correlation functions are derived, and there is some discussion of
the issues involved with the continuum limit.

\end{abstract}
\section{Introduction}
Nonrenormalizable quantum field theories have long been the ``black
sheep of the family''.\footnote{Common meaning: ``A worthless or
disgraced member of the family''. See, for example,
http://www.phrases.org.uk/meanings/66250.html.} While
super-renormalizable and strictly renormalizable quantum field
theories (especially asymptotically-free theories) have been well
served by renormalized perturbation theory, this has not been the
case for the usual nonrenormalizable family members. For them -- so
the perturbation story goes -- an infinite set of distinct
counter-terms are necessary, requiring thereby an infinite number of
experiments to establish the needed coefficients of the
counterterms, an endeavor which nullifies any predictive power the
theory may have had. A key concept in the preceding sentence relates
to the perturbative counterterms, and we shall stress the
shortcomings of this concept below.

Nonrenormalizable models may also be approached nonperturbatively.
Here, we have in mind the analysis of $\varphi^4_n$ models for
spacetime dimension $n\ge5$ made by Aizenman \cite{aiz} and
Fr\"ohlich \cite{fro}. In their approach, they formulated the models
as Euclidean-space lattice theories with arbitrary coefficients for
the mass term and the coupling constant, both taken as functions of
the cutoff parameter, namely, the lattice spacing $a$; of course,
the coupling constant was required to be nonnegative. They were
rigorously able to show, independent of the choice of the indicated
parameters, that the continuum limit (as combined with the infinite
volume limit) of the theory led to the result of a (possibly
generalized) free field. In other words, the manifestly non-Gaussian
nature of the lattice field distribution has, in the continuum
limit, become a strictly Gaussian distribution.  Since Monte Carlo
calculations suggest that triviality holds for $\varphi^4_4$ as
well, this overall story has given rise to the widely held view of
the ``triviality of the $\varphi^4$ theories''.

In the author's judgement, the infinite number of distinct counter-
terms as well as the triviality result, while correct conclusions
given their respective underlying assumptions, are both unacceptable
as physically correct quantum statements for the problem at hand.
This conclusion is based on the following assertions: (1) For a
nonrenormalizable theory, results based on regularization and the
introduction of counterterms given by a perturbation analysis about
the free theory are unacceptable for the simple reason, as we shall
argue, that the interacting nonrenormalizable theories are {\it not
even continuously connected} to the free theory as the coupling
constant is reduced to zero; (2) The {\it classical} $\varphi^4_n$,
$n\ge5$, models are manifestly {\it non}trivial theories exhibiting
nonvanishing scattering, etc. If the quantum formulation of such
models is taken to be the trivial one, then it follows that the
classical limit of that quantum theory leads to a trivial classical
theory which is {\it not} equivalent to the original classical
theory one started with. Such a dilemma can only mean that the
quantum formulation leading to triviality cannot be the physically
correct quantum formulation for nonrenormalizable $\varphi^4_n$,
$n\ge5$, models.

We claim there is an alternative formulation of the quantum theory
for such models that can explain the unacceptable results of
infinitely many counterterms as well as triviality, and which
furthermore offers alternative calculational possibilities that we
believe may lead to satisfactory results. The purpose of the present
paper is to lay out the theoretical arguments supporting the point
of view under present consideration. We start by reexamining the
limiting Gaussian behavior of the $\varphi^4_n$, $n\ge5$, models.

In probability theory, the Central Limit Theorem (CLT) leads to
Gaussian distributions for a very large class of independent,
identically distributed random variables. By concentrating attention
on random variables that are independent and identically
distributed, it is of course possible to demonstrate the CLT quite
generally. However, it stands to reason that there exist
distributions of random variables that are neither identical nor
independent, but which are close to such behavior in some
unspecified manner, yet the combination nevertheless tends to a
Gaussian distribution as the number of variables increases without
limit. General theorems for such cases are naturally most unlikely
since a Gaussian or non-Gaussian limit depends explicitly on the
particular details at hand. Nevertheless, based on the evidence, it
is plausible to regard $\varphi^4_n$, $n\ge5$, models as satisfying
the criteria to lie within the
 basin of attraction of the CLT. Our further development will be
 partially influenced by this viewpoint.

 In Sec.~2 we outline the argument why nonrenormalizable
 interactions act as discontinuous perturbations, and as such,
 counterterms suggested by regularized and renormalized perturbation
 theory are unsuitable. Attention shifts from the lattice action and the field
 distribution it engenders in Sec.~3, and instead focusses on the
 sharp-time, ground-state distribution. Arguing from the CLT, we
 propose that this distribution should be a generalized Poisson
 distribution, a subject discussed at length in
 Sec.~4. A generalization to a linear superposition of
 Poisson distributions is the subject of Sec.~5. It
 is shown in Sec.~6 that general spacetime correlation functions can be
 expressed as, or are at least bounded by, properties of the Poisson
 distribution, while Sec.~7 offers a useful reformulation of some of the
 expressions developed in Sec.~6 as well as a brief discussion of the
 continuum limit.  Section 8 is devoted to a
 conclusion, and some details about specific properties of
 a given generalized Poisson distribution are discussed in the Appendix.

 It is not without interest that in an archived (but unpublished)
 paper \cite{klau}, the author already made the proposal that the ground-state distribution
 for nonrenormalizable scalar models should be a generalized Poisson
 distribution. Unfortunately, how that proposal was used in that earlier paper to suggest
 an auxiliary potential and thus a lattice action was incorrect.
 Hopefully that situation has been advanced in the present paper.

 \section{Inadequacy of Regularized, Renormalized Perturbation Theory}
 \subsection{The hard-core picture}
 Chapter 8 in \cite{book} describes in some detail the hard-core
 picture of nonrenormalizable interactions, and for a detailed
 account of that approach we can do no better than to encourage the
 reader to examine that presentation, which, importantly, is for all
 intents and purposes an independently readable chapter. However, for
 the sake of the reader who does not need a detailed account of the
 hard-core picture of nonrenormalizability, we offer the following
 brief synopsis.

 In general terms, let the theory in question be described by a
 formal (Euclidean-space) functional integral commonly called a
 generating functional, which for the ``free'' theory is given by
  \b S_0\{h\}\equiv{\cal N}_0\int e^{\tint h\s\phi\s
  d^n\!x\s-\s W_0(\phi)}\;{\cal D}\phi\;,  \e
  where $h$ is a smooth ``source'' function, and $W_0(\p)\ge0$ denotes
  the free field (Euclidean) action,
  an expression which is often quadratic in the fields but need not
  be so. Likewise we have the generating functional for the ``interacting'' theory
   \b S_\l\{h\}\equiv{\cal N}_\l\int e^{\tint h\s\p\s d^n\!x\s-\s W_0(\p)-\l\s
   V(\p)}\;{\cal D}\p\;,  \e
   where the coupling constant $\l>0$, and $V(\p)\ge0$ denotes the
   (Euclidean) interaction action. In each case, the formal normalization is
   chosen so that $S_0\{0\}=S_\l\{0\}=1$.

   We say that $V$ is a {\it
   continuous} perturbation of $W_0$ if
     \b \lim_{\l\ra0}\;S_\l\{h\}=S_0\{h\}\;,  \e
   for all $h$ (say $C^\infty_0$); in addition, we say that $V$ is a {\it discontinuous}
   perturbation of $W_0$ if, instead,
      \b \lim_{\l\ra0}\;S_\l\{h\}=S'_0\{h\}\ne S_0\{h\}\;, \e
   for all $h$, where $S'_0\{h\}$ is referred to as the
   generating functional for the ``pseudofree'' theory.

   At first glance, it seems self evident from the given expressions for the
   generating functionals that the limit of
   $S_\l\{h\}$ as $\l\ra0$ must be $S_0\{h\}$. Moreover, the unquestioned
   belief that this must be so would appear to be a tacit
assumption behind any
   perturbative analysis of $S_\l\{h\}$ in a series
   expansion in $\l$ about $\l=0$. However formal (i.e., nonconvergent) that
   series may be, the introduction and use of such a series expansion
   presumes that $V$ is a continuous perturbation of $W_0$. If $V$ is
   {\it not} a continuous perturbation, i.e., if $V$ is in fact a {\it
   discontinuous} perturbation of $W_0$, then there is no way in
   which one could legitimately claim that ``$S_\l\{h\}$ has a
   perturbation series expansion about $S_0\{h\}$''. That
   statement is as likely to be true as the statement that
   properties of the (discontinuous!) function
   \b &&F(\l)\equiv e^\l,\hskip.6cm \l>0\;, \\
       &&F(0)\equiv \pi                     \e
   can be determined by a power series expansion about $\l=0$.

   Qualitatively speaking, a continuous perturbation is one for which
     \b  W_0(\p)<\infty \hskip.3cm \Longrightarrow \hskip.3cm V(\p)< \infty\;.  \e
   Similarly, a discontinuous perturbation is one for which
      \b W_0(\p)<\infty \hskip.3cm \not\!\!\Longrightarrow \hskip.3cm V(\p)< \infty\;, \e
   or in other words, there are fields (comprising a set of nonzero measure) for which
    \b W_0(\p)<\infty\;,\hskip.7cm V(\p)=\infty\;.  \e
   For a discontinuous perturbation, the contributions such fields would have made if the interaction
   term $V$ had never been present are now missing altogether because those fields
   are effectively projected out by the factor $e^{-\l \s V}$, for any $\l>0$, however small.
   If we introduce
     \b  && X(\p)\equiv1 \;,   \hskip.4cm {\rm if}\hskip .6cm W(\p)<\infty\;, \hskip.3cm V(\p)<\infty\;,  \\
           && X(\p)\equiv0 \;,   \hskip.4cm {\rm if}\hskip .6cm W(\p)<\infty\;, \hskip.3cm V(\p)=\infty\;, \e
   then we see that we can freely redefine $S_\l\{h\}$ as
    \b S_\l\{h\}={\cal N}_\l\int e^{\tint h\s\p\s d^n\!x\s-\s W_0(\p)\s-\s\l\s V(\p)}\,X(\p)\,{\cal D}\p\;. \e
    Now, as $\l\ra0$, we can readily see that we obtain
     \b S'_0\{h\}={\cal N'}_0\int e^{\tint h\s\p\s d^n\!x-W_0(\p)}\,X(\p)\,{\cal D}\p\;.  \e
In brief, the interaction term has acted partially as a hard core
projecting out -- thanks to $X(\p)$ -- certain contributions that
would have been included for the free generating functional!

To show the relevance of this discussion to $\p^4_n$ models, we need
only recall for $n\le4$ (the renormalizable cases) that
 \b  \{\s\tint \p^4(x)\,d^n\!x\s\}^{1/2}\le (4/3)
 \s\tint\s\{\s[\nabla\s\p(x)\s]^2+m^2\s\p^2(x)\s\}\,d^n\!x \e
 for any $m^2>0$, while for $n\ge5$ (the nonrenormalizable cases)
 there are singular fields, such as
   \b  \p_s(x)=\frac{e^{-x^2}}{|x|^p}\;, \hskip.7cm \frac{n}{4}<p<\frac{n}{2}-1\;, \e
   for which
   \b  \tint\s\{\s[\nabla\s\p_s(x)\s]^2+m^2\s\p_s^2(x)\s\}\,d^n\!x <\infty\;, \e
   while
    \b  \tint \p_s^4(x)\,d^n\!x=\infty \;,\e
    fulfilling the requirement for the interaction to act as a hard core.

    A derivation of the $(4/3)$ bound given above when $n\le4$  is part of the discussion
    in Chapter 8 of \cite{book}.
    The factor $(4/3)$ also provides a bound for the more general $\varphi^p_n$
    models when $p\le2n/(n-2)$ (the renormalizable cases), while
    there is no finite bound when $p>2n/(n-2)$ (the nonrenormalizable cases).
    When the gravitational action is
    divided into quadratic and nonquadratic terms, the latter term has the
    properties to act as a discontinuous
    potential consistent with the well-known nonrenormalizability of gravity.
    In fact, as also detailed in Chapter 8
    of \cite{book},
    discontinuous perturbations of the harmonic oscillator arise even in
    conventional quantum mechanics.
    If hard-core potentials
    exist in quantum mechanics, it should be no surprise that they arise
    in quantum field theory!

We are about to leave the general description of hard-core
interactions hoping that the reader has accepted that hard-core
interactions may indeed exist and that nonrenormalizable quantum
field models are likely candidates. Harder to accept, but
nevertheless a direct consequence of the existence of hard-core
interactions, is the concept that {\it in such cases} counterterms
suggested by a (regularized) perturbative power series expansion
about the original {\it free} theory are not relevant -- indeed,
harsh as it may seem, we can even say they are {\it completely
irrelevant!} In fact, such terms are {\it entirely misleading}
because regularized, perturbative counterterms are {\it designed} to
preserve a continuity of the interacting theory with the original
free theory, a continuity which in fact {\it never existed!}

This statement about perturbation theory is no doubt difficult to
accept for some readers because it certainly goes against -- even
contradicts -- the ``general wisdom'' of how nonrenormalizable
models are normally presented. If the reader's prejudice in favor of
perturbation theory and its unquestioned, universal applicability
prevent you from accepting the statement that in genuine hard-core
cases perturbation theory about the original free theory is
irrelevant, then this article is not for you. By all means, stop
reading here and now! However, if you can accept the fact that
hard-core interactions have the power to change almost everything
regarding perturbation theory about the free theory, then we invite
you to read on. Indeed, we can offer you the comforting possibility
that the interacting theory may indeed have a meaningful
perturbation expansion -- not about the free theory -- but about the
{\it pseudofree} theory, which after all is the theory to which it
is continuously connected as $\l\ra0$! As an elementary illustration
of that very thought, let us revisit the discontinuous function
$F(\l)$ introduced earlier. This time, however, we introduce an
analogue of the pseudofree theory by letting
  \b &&F'(\l)\equiv F(\l)=e^\l\;,\hskip.6cm \l>0\;,   \\
     &&F'(0)\equiv \lim_{\l\ra0}F(\l)=1\;.  \e
Clearly -- and unlike the original function $F(\l)$ -- the new
function $F'(\l)$ does have a perturbative power series
representation about the ``pseudofree'' value, $F'(0)=1$, namely,
  \b F'(\l)=\sum_{n=0}^\infty\frac{\l^n}{n!}\;, \e
which is valid in the present case for all $\l\ge0$. More
importantly, highly specialized, soluble nonrenormalizable models
(see Chapters 9 \& 10 in \cite{book}) possess solutions that {\it
do} admit entirely reasonable perturbative treatments of their
interactions about the appropriate pseudofree solution. We believe
valid perturbative analyses about suitable pseudofree solutions hold
widely, and we will make the assumption in this paper that such
perturbation formulations exist.

Accepting the concept that the nonlinear interaction for
$\varphi^4_n$, $n\ge5$, models acts partially like a hard core leads
immediately to the conclusion that some form of counterterm is
needed, at the very least, to represent the irremovable effects of
the hard core. The triviality argument also supports the need for
counterterms since mass, coupling constant, and field-strength
renormalization alone are insufficient to escape triviality. Our
previous discussion has been necessary to open the way to consider
counterterms other than those suggested by renormalized perturbation
theory. Our task is to find suitable counterterms that will avoid
triviality -- a highly nontrivial task to be sure!

\section{Focus on the Sharp-Time, Ground-State \\Distribution}
\subsection{Triviality also holds at sharp time}
 A Euclidean spacetime
distribution that exhibits triviality is necessarily Gauss- ian, and
has vanishing truncated correlation functions beyond second order.
Such correlation functions {\it a fortiori} vanish at equal times, a
fact that implies that the associated ground-state distribution for
the limiting Hamiltonian has a Gaussian distribution itself. Stated
otherwise, the lattice-space ground-state distribution -- the square
of the lattice-space ground-state wave function itself -- which
because of the nonlinear interaction is manifestly {\it
non}-Gaussian, has, in the continuum limit, become Gaussian. Just as
the full-time distribution has become Gaussian because it lies in
the appropriate basin of attraction for the CLT, it is equally
plausible to conclude that the sharp-time field distribution for the
nonlinear theory also lies in the associated basin of attraction of
the CLT. Conversely, if we can ensure that the ground-state
distribution remains {\it non}-Gaussian, then the non-Gaussian
character of the full-time distribution will be assured. This remark
leads us to focus some attention on the putative ground state.

\subsection{Lattice space version of the ground-state wave function}
We let the even, positive function $\Psi(\p)\;[=\Psi(-\p)]$, denote
the ground-state wave function on the Euclidean-space lattice, and
we adjust the potential by a constant so that this ground state has
zero energy. Consequently, the nonnegative Hamiltonian operator for
the lattice is given by
   \b  \H = -\frac{1}{2}\, a^{-s}\s{\sum_k}'\frac{\d^2}{\d \phi_k^2}+{\cal V}(\p)\;,  \e
where the potential is given by
   \b {\cal V}(\p)\equiv\frac{1}{2}\, a^{-s}
   \s{\sum_k}'\frac{1}{\Psi(\p)}\s\frac{\d^2\s\Psi(\p)}{\d\phi_k^2}\;. \e
Let us assume that the lattice has lattice spacing $a$,
  is hypercubic with periodic boundary conditions in all coordinate-axis directions, and the
  number of lattice sites $k=(k_0,k_1,k_2,\ldots,k_s)$, $k_j\in {\mathbb Z}$, $s=n-1$,
  in each axial direction is $L$.
Here the primed sum $\S_k$ signifies a sum over spatial values of
$k$, i.e.,
   $k_1,k_2,\ldots,k_s$, at some specific, but generally implicit, fixed, temporal value of $k$, i.e., $k_0$.
The argument $\p$ stands for the collection of lattice fields
$\{\p_k\}$ at an equal lattice time. Thus if
$k=(k_0,k_1,k_2,\ldots,k_s)$, we select $k_0$ as the future time
direction (under a possible Wick rotation), and so $\p=\{\p_k:\;{\rm
all}\; k \;{\rm with}\; k_0\;{\it \rm fixed}\}$.

The probability distribution constructed as the square of the
ground-state wave function may alternatively be characterized by its
characteristic function
  \b C(h)\equiv\int e^{i\s\Sigma'_k\s h_k\s\p_k\,a^s}\,\Psi(\p)^2\,\Pi'_k\s d\p_k\;; \e
the symbol $\Pi'_k$ denotes a product over spatial values of $k$
holding the temporal value fixed. We define the continuum limit of
such expressions to mean: (i) the lattice spacing $a\ra0$ as well as
(ii) the spatial volume $(La)^s\ra\infty$ along with the spacetime
volume $(La)^n\ra\infty$ in an appropriate manner and combination.
As noted above, if we choose the naive lattice form (including
arbitrary cutoff dependent parameters) for the classical action of a
$\varphi^4_n$, $n\ge5$, model, the ground-state distribution becomes
Gaussian, or equivalently that
  \b C(h)\ra E_{Gaussian}(h)
\equiv\exp[-\half\tint h(x)\s U(x-y) \s h(y)\,d^s\!x\,d^s\!y]\;,  \e
for some covariance function $U(x-y)$. Our primary goal is to {\it
avoid} a Gaussian limiting behavior in the continuum limit for the
ground-state distribution.

\subsection{Some partial matrix elements}
The lattice-space Hamiltonian discussed above admits an alternative
and useful form. Let us introduce the differential operator
  \b A_k\equiv -\,a^{-s}\s\frac{\d}{\d\p_k}+a^{-s}\s\frac{1}{\Psi(\p)}\s\frac{\d
  \Psi(\p)}{\d\p_k}\;, \e
  and its adjoint operator
  \b A^\dagger_k\equiv \,a^{-s}\s\frac{\d}{\d\p_k}+a^{-s}\s\frac{1}{\Psi(\p)}\s\frac{\d
  \Psi(\p)}{\d\p_k}\;, \e
for each $k$ in a given spatial lattice slice. It is straightforward
to verify that the lattice Hamiltonian is also given by
  \b  \H=\half\s\S_k\s A^\dagger_k\s A_k\s a^s\;, \e

  Next consider states of the form
    \b \Psi_h(\p)\equiv e^{i\S_k h_k\s\p_k\s a^s}\,\Psi(\p)\;, \e
    and note that
    \b  A_k\s\Psi_h(\p)=-\s i\s h_k\s\Psi_h(\p)\;.  \e
  Consequently, it follows that
   \b \int \Psi_{h'}(\p)^*\s\H\s\Psi_h(\p)\,\Pi'_k\s
   d\p_k=\half\s\S_k\,
   h'_k\s h_k\, a^s\,C(h-h')\;.   \e
   Additionally, we observe that
     \b  -\frac{i}{2}(A_l^\dagger\s-\s A_l)=-i\s\s
     a^{-s}\s\frac{\d}{\d\p_l}\equiv \pi_l\;,\e
     the canonical momentum conjugate to the field $\p_l$. It
     follows that
        \b \int\Psi_{h'}(\p)^*\s\pi_l\s\Psi_h(\p)\,\Pi'_k\s d\p_k=\half(h'_l+h_l)\,C(h-h')\;. \e
 Equivalently, if we use the connection
    \b \pi_l=-i\s a^{-s}\frac{\d}{\d\d_l}=i[\H\s,\s\p_l]\;  \e
relating the canonical momentum with the field and Hamiltonian, then
it follows that
  \b &&\int \Psi_{h'}(\p)^*\s\pi_l\s\Psi_h(\p)\,\Pi'_k\s
  d\p_k=a^{-s}\s\bigg(\frac{\d}{\d h'_k}+\frac{\d}{\d h_k}\bigg)\,
  \int\Psi_{h'}(\p)^*\s\H\s\Psi_h(\p)\,\Pi'_k\s d\p_k\\
   &&\hskip4.67cm =\half(h'_l+h_l)\,C(h-h')\;, \e
   as before. ({\bf Remark:}
There are now THREE different uses of the prime: one refers to the
pseudofree theory; another refers to properties on a fixed time
slice of the lattice; the last usage refers to labels of Hilbert
space functions. The context suffices to tell which usage is meant.)

   Observe that this calculation has yielded matrix elements of both
   the lattice Hamiltonian $\H$ and the lattice momentum $\pi_l$ at
   site $l$. It is highly probable that these matrix elements
   actually determine the operators in question, albeit in an
   indirect manner.

\subsection{Features of the continuum theory}
The analysis given above for the lattice has a natural analog in the
continuum theory \cite{araki}. The expressions given below should
indeed follow directly as the continuum limit of the lattice
expressions, despite the fact that we have changed notation to
emphasize the fact that we are dealing with the putative continuum
theory.

 The continuum limit of the ground state may be
formally identified with the abstract unit vector $|\s0\>$, a member
of the abstract Hilbert space ${\frak H}$. Additionally, we can
introduce the unit vectors
  \b |h\>\equiv e^{i\tint h(x)\s\hph(x)\,d^s\!x}\,|\s0\>\;, \e
where $s=n-1$, $\hph(x)$ denotes the sharp-time (e.g., at $t=0$),
formally self-adjoint field operator, and $h(x)$ denotes a smooth,
real test function. Additionally, the vector $|h\>$ is strongly
continuous in a suitable topology \cite{hegkla}. The overlap of two
such vectors,
  \b E(h-h')\equiv \<h'|h\> \;, \e
defines the important expectation functional $E(h)$. As a continuous
function of positive type, $E(h)$ serves as a reproducing kernel for
a reproducing kernel Hilbert space that provides a representation of
${\frak H}$ by continuous functionals. Furthermore, if we define the
local operator $\hp(x)$ as the canonical conjugate of the field
operator $\hph(x)$, then, under mild conditions, it follows
\cite{araki} that
  \b \<h'|\hp(x)|h\>=\half[h'(x)+h(x)]\,E(h-h')\;.  \e
Since we may assume that the vectors $|h\>$ span the Hilbert space,
it is plausible that the given matrix elements of $\hp(x)$ determine
the local self-adjoint operator uniquely. Moreover, we can also
assert \cite{araki} that
  \b \<h'|\s\H\s|h\>=\half (h',h)\,E(h-h')\;,  \e
where $\H$ denotes the nonnegative, self-adjoint Hamiltonian
operator for which $\H\s|\s0\>=0$, provided we assume that
  \b \hp(x)=i[\H,\hph(x)]\;.  \e

  Note well, that these relations which hold in the continuum are
  simply natural continuum analogs of the corresponding lattice-space
  expressions derived above, and confirm that a smooth continuum limit of the
  function $C(h)$ to $E(h)$ is a primary ingredient in developing
  a satisfactory continuum theory.


\section{The Role of Generalized Poisson \\Distributions} In probability theory,
alternative distributions of the sum of an ever-increasing number of
independent, identically distributed random variables that compete
successfully in the struggle for convergence, such as in a continuum
limit, generally lead to Poisson distributions, and in doing so they
give rise to non-Gaussian final distributions. Significantly, a
sequence of lattice-space ground state distributions with a limiting
behavior that leads to a Poisson distribution necessarily has a
qualitatively different dependence on the individual variables and
various parameters than does a sequence of lattice-space ground
state distributions destined to end up as a Gaussian distribution.
Just as for the Gaussian limit, it is plausible that there are
distributions for alternative sets of random variables, which are
neither independent nor identically distributed, yet lie in suitable
basins of attraction so that they have limiting Poisson
distributions. If we ensure that the terms in the lattice action
along with appropriate counterterms conform with the needed
different dependence on the individual variables and parameters,
then we can ensure that the continuum limit will avoid being
Gaussian.

As stressed previously, the traditional free theory is Gaussian and
therefore corresponds to an infinitely divisible distribution
(defined below). There is some logic in the fact that a free theory
is infinitely divisible in the sense that the ability that each
field variable may be expressed as a sum of arbitrarily many
independent, identically distributed variables is, let us say, one
hallmark of being ``free''. Likewise, when discussing a pseudofree
theory in which all interactions have been reduced to zero, it is
not unrealistic to believe that the hallmark of infinite
divisibility also applies to the resultant pseudofree distribution.
While arguments of any weight are hard to advance, we shall
nevertheless make the assumption that it is possible that this is
the case and, as a consequence, we shall seek pseudofree solutions
among the family of infinitely divisible distributions. Even so, we
shall also hedge our bets by suggesting that the class of
ground-state distributions which are given by rather general linear
combinations of infinitely divisible distributions (and are
therefore, in general, {\it not} infinitely divisible themselves)
may be the proper place to find pseudofree theories.

Before studying particular examples of infinitely divisible
distributions, however, it is appropriate to give a general outline
of how such distributions may be characterized.

Successful limiting behavior of suitable sets of countable numbers
of independent, identically distributed random variables may be
formalized in a natural way in terms of their characteristic
functions. In particular, the appropriate Gaussian and Poisson
distributions share the property of {\it infinite divisibility},
which may be described as follows: If $C(h)$ denotes an infinitely
divisible characteristic function for an appropriate distribution,
then it follows that the $J^{\rm th}$ root of that function, i.e.,
$C(h)^{1/J}$, $J\in\s\{2,3,4,\ldots\}$, is also a characteristic
function. This property follows because we can always linearly
decompose the set of random variables $\{\p_k\}$ into $J$ components
composed of independent, identically distributed random variables
for any positive integer $J$. Furthermore, it is always the case
that any integer power $K$ of a characteristic function is again a
characteristic function, i.e., if $C(h)$ is a characteristic
function, then $C(h)^K$, $K\in\{2,3,4,\ldots\}$ is also a
characteristic function. Therefore, an infinitely divisible
characteristic function $C(h)$ retains the property of being a
characteristic function if we raise it to an arbitrary rational
power such as $C(h)^{K/J}$. Finally, since characteristic functions
are always continuous functions of their arguments, we can take a
limit as the rational ratios converge to an arbitrary, positive
real, $K/J\ra r$. This limit converges, and the limit $C(h)^r$ is
again a characteristic function for an arbitrary real number $r>0$.
This feature is a defining as well as a useful property of
infinitely divisible characteristic functions and thereby of
infinitely divisible distributions.

The only categories of infinitely divisible distributions are either
Gaussian or Poisson, or a combination of the two. And since a
Gaussian distribution can be obtained as the limit of a sequence of
Poisson distributions (but not vice versa!) it is appropriate to say
that all infinitely divisible distributions are Poisson
distributions or limits thereof.

In what follows we shall assume that the ground-state distribution
function has the form of a fairly straightforward, basic Poisson
distribution. We conjecture that some of the basic Poisson
distributions on which we focus could possibly serve as appropriate
pseudofree scalar field models.

In Sec.~5, ground-state distributions that are given by linear
superpositions of basic Poisson distributions are treated as
possibly interesting generalizations of the case of simple Poisson
distributions.

In seeking models within the class of strictly infinitely divisible
distributions, or within the class of linear superpositions of such
distributions, it is important to keep in mind that some {\it
specific} model, such as $\varphi^4_n$, $n\ge5$, may {\it not} be
among those contained in the class under consideration. Indeed, one
should not look for {\it any specific} model, but, at the present
stage,  accept any reasonable model that might be present. This view
is necessitated because we are faced with a highly nontrivial {\it
inverse problem}; namely, we can access a limited amount of
information about the model under consideration, i.e., some features
of the ground-state distribution, and from that we would like to
determine the lattice action that gave rise to those properties. In
summary, we ask the reader to keep an open mind as to just what
class of models of interest -- if any -- can be described by the
procedures we have in mind.

\subsection{Basic Poisson distributions}
To describe a Poisson
distribution for scalar field models in terms of the ground-state
distribution $\Psi(\p)^2$ is difficult because it generally does not
involve known functions. On the other hand, the characteristic
function for a general Poisson distribution can be described in a
relatively simple form. In particular, the characteristic function
of an even Poisson distribution has the form given by
  \b \int \cos(\S_k h_k\p_k\s a^s)\,\Psi(\p)^2\;\P_k\s d\p_k
  =\s \exp\{\s-\tint[1-\cos(\S_kh_k\p_k\s a^s)]\s\rho(\p,a)\,\P_k\s d\p_k\s\}\;, \e
where the weight function $\rho(\p,a)\ge0$.

  To ensure absolute continuity of the ground-state distribution, it is necessary and sufficient
  that
     \b \tint \rho(\p,a)\,\P_k\s d\p_k = \infty \;, \e
     a criterion that establishes our choice of Poisson
     distributions as so-called generalized Poisson distributions \cite{defin}.
     And to ensure proper meaning of the right-hand side of the equation
     for the generalized Poisson characteristic function above, it is necessary that
the exponent -- frequently referred to as the ``second
characteristic'' \cite{luk} -- satisfies
   \b  0\le\tint[1-\cos(\S_kh_k\p_k\s a^s)]\s\rho(\p,a)\,
\P_k\s d\p_k\s<\infty  \e for all suitable $\{h_k\}$.
 By symmetry, the weight function $\rho(-\p,a)=\rho(\p,a)$, and
we assume that all
 even ``moments'' exist in the sense that
   \b \tint \p_{k^1}\p_{k^2}\cdots\p_{k^{2q}}\,\rho(\p,a)\,\P_k\s
d\p_k<\infty\;, \e for all $q\ge1$, where
$k^m\equiv(k_0,k^m_1,k^m_2,\ldots k^m_s)$, and the time $k_0$ is the
same for all $k^m$, $1\le m\le 2q$. By assumption all odd
``moments'' vanish. Note that such expressions are {\it not} moments
since the weight function $\rho(\p,a)$ is not -- and cannot be --
normalized. To emphasize that distinction we shall generally refer
to such expressions as {\it noments} (rhymes with moments, and is
derived from {\it n}ot m{\it oments}, a phrase that determines the
meaning of noments).

In order that $\rho(\p,a)$ have the indicated properties, it must be
singular near $\p=0$ in a suitable manner. For our purposes we shall
focus on $\rho(\p,a)$ of the form
  \b \rho(\p,a)\equiv \frac{R(a)\s e^{-U(\p,a)}}{\P_k[\S_l
\s\beta_{k-l}\s\p^2_l]^\gamma}\;. \e Let us explain the separate
terms that enter this expression.

  The numerator contains a factor $R(a)$ to be fixed later and
an unspecified function
  $U(\p,a)$, which is limited at present, let us say, by the requirements that $U(0,a)=0$ and
  $U(\p,a)\ge c\s\S_k\p^2_ka^s$, for some $c>0$, and which is defined for field variables
  at a fixed time $k_0$; at the present
  moment, the only purpose of $U(\p,a)$ is to ensure that all noments of $\rho(\p,a)$ are finite.
  Below, we shall further specialize $U(\p,a)$ to the quadratic form
   \b U(\p,a)=\S_{k,l}\p_k\s A_{k,l}\s\p_l\s a^{2s}\;, \e
   for suitable (but unspecified) matrices $A_{k,l}$ for closer examination as candidates for
   pseudofree models; we shall even take the liberty of calling models with this form
   ``pseudofree models''. As will become clear, one argument for this designation
   is the fact that in its final form, this particular version of the weight function
   $\rho(\p,a)$ has the same number of free, dimensional parameters as a
   traditional free theory. Additional arguments in favor of such a
   designation will arise later.

  As shown in the Appendix, the exponent $\g$ in the denominator of the
  expression for $\rho(\p,a)$ satisfies the bounds
  \b\frac{1}{2}\le \g<\frac{1}{2}+\frac{1}{\s N'}\;, \e
  where $N'=L^s=L^{n-1}$
 denotes the number of lattice sites in a spatial slice. Since we eventually
need to take $N'\ra\infty$, we accommodate that limit already by
choosing
  \b  \g=\frac{1}{2}\;,   \e
   the only exception being in the Appendix where
  we establishing the bounds on $\g$ noted above. Since the
  parameters $\{\beta_k\}$ will turn out to be dimensionless, then, when
  $\g=1/2$, it follows that $R(a)$ is dimensionless. Consequently, all the
  dimensional parameters in $\rho(\p,a)$ reside in $U(\p,a)$.

The factors
  $\B_k\,[\s=\B_{-k}]$ are all nonnegative, i.e., $\B_k\ge0$, and satisfy the condition that
    \b  \S_k\s\B_k=1\;. \e
Various choices of the set $\{\B_k\}$ are possible, but we shall
concentrate on choosing a fixed, $N'$-independent number of $\B_k$
terms as nonzero. In particular, we specialize to the specific
choice of $2s+1$ nonvanishing terms given by
    \b \B_k=J_k\equiv\frac{1}{2s+1}\s\delta_{\s k,\{k\s\cup \s k_{nn}\}}\;. \e
This notation means that an equal weight of $1/(2s+1)$ is given to
the $2s+1$ points in the set composed of $k$ and its $2s$ nearest
neighbors in the spatial sense only. Specifically, we define
$J_k=1/(2s+1)$ for the points $k=(k_0,k_1,k_2,\ldots,k_s)$,
$k=(k_0,k_1\pm1,k_2,\ldots,k_s)$,
$k=(k_0,k_1,k_2\pm1,\ldots,k_s)$,\ldots,
$k=(k_0,k_1,k_2,\ldots,k_s\pm1)$, and $J_k=0$ for all other points
in that spatial slice.\footnote{Originally \cite{klau}, it was
believed that it was necessary that all $\B_k>0$ was required. We
thank Erik Deumens for raising the question of whether finitely many
positive $\B_k$ terms might work just as well as requiring that all
$\B_k>0$. Happily, that has turned out to be the case.} The
suitability of this choice is addressed in the Appendix. For
notational purposes we shall refer to the set of lattice sites
   $\{\s k\s\cup\s k_{nn}\s\} $
   as the set ``$\kk$''. The name $\kk$ is chosen because for the graphically simple case for which
   $s=2$,
the set of nonvanishing $\{J_k\}$ points are arranged in the form of
a ``$+$'' sign; indeed, see Fig.~1 in the Appendix  in this regard.
  ({\bf Remark:} Of course, one could give an unequal, positive weighting to each of the $\kk$ points
   in contrast to the set of equal weights we have chosen. While this choice would lead to a
   different sequence of results for finite lattice spacings, any  difference due to unequal
   weighting would disappear in the continuum limit. Such an argument applies to any set $\{\B_k\}$
   that contracts to a single point in the continuum limit.)

  Clearly, not all ground-state distributions $\Psi(\p)^2$ correspond
  to infinitely divisible distributions. However, for every
  acceptable choice of $U(\p,a)$, the prescription that defines $C(h)$ {\it automatically} defines --
  albeit indirectly -- a ground-state distribution $\Psi(\p)^2$ that is
  infinitely divisible.

\subsection{What is the pseudofree ground-state distribution?}
Given a characteristic function for a generalized Poisson measure,
the associated distribution is given be an inverse Fourier
transform, namely,
 \b
&&\hskip-.7cm\Psi(\p)^2\equiv\bigg(\frac{a^s}{2\pi}\bigg)^{N'}\int\cos(\S_k\p_kh_k\s
  a^s)\\
  &&\hskip0cm\times\exp\{-\s\tint[1-\cos(\S_kh_k\s u_k \s a^s)]\,\rho(u,a)\,\Pi'_k\s
  du_k\}\,\Pi'_k dh_k\;.  \e
In general, this integral cannot be evaluated analytically; however,
there is a partial result which is illuminating. It follows from the
property of infinite divisibility that $C(h)^{1/M}$ is again a
characteristic function for all $M\in\{2,3,4,\ldots\}$, and we can
put that property to use for us now.

Consider the expression
  \b &&\hskip-.7cm\Psi_{1/M}^2(\p)\equiv\bigg(\frac{a^s}{2\pi}\bigg)^{N'}\int\cos(\S_k\p_kh_k\s
  a^s)\\
  &&\hskip0cm\times\exp\{-(1/M)\s\tint[1-\cos(\S_kh_k\s u_k \s a^s)]\,\rho(u,a)\,\Pi'_k\s
  du_k\}\,\Pi'_k dh_k\;,  \e
  which yields the distribution belonging to the characteristic
  function $C(h)^{1/M}$. For very large $M$ it stands to reason that
  the second characteristic is small, which allows us to approximate
  the foregoing integral up to order $O(1/M^2)$ as follows:
  \b &&\hskip-.7cm\Psi_{1/M}^2(\p)\equiv\bigg(\frac{a^s}{2\pi}\bigg)^{N'}\int\cos(\S_k\p_kh_k\s
  a^s)\\
  &&\hskip0cm\times\bigg[\s1-(1/M)\s\tint[1-\cos(\S_kh_k\s u_k \s a^s)]\,\rho(u,a)\,\Pi'_k\s
  du_k\,\bigg]\,\Pi'_k dh_k\;,  \e
  Now we make use of the specifically assumed form for $\rho(\p,a)$ given by
    \b \rho(\p,a)=\frac{R(a)\s\exp[\s-\s U(\p,a)\s]}{\Pi'_k[\S_lJ_{k-l}\s
    \p_l^2]^{1/2}}\;, \e
    and remark, since we are working to order $O(1/M)$ that we can
    replace this weight function by (say)
     \b \rho_{1/M}(\p,a)\equiv\frac{R(a)\s\exp[\s-\s U(\p,a)\s]}{\Pi'_k[\S_lJ_{k-l}\s
    \p_l^2\s+\s F(M)]^{1/2}}\;, \e
where the additional term in the denominator, $F(M)$, satisfies the
bounds
   \b 0<F(M)\ll 1\;. \e
   With this replacement in place, we now claim, with the same
   accuracy as before, that
\b
&&\hskip-.7cm\Psi_{1/M}^2(\p)=\bigg(\frac{a^s}{2\pi}\bigg)^{N'}\int\cos(\S_k\p_kh_k\s
  a^s)\\
  &&\hskip0cm\times\bigg[\s1-(1/M)\s\tint[1-\cos(\S_kh_k\s u_k \s a^s)]\,\rho_{1/M}(u,a)\,\Pi'_k\s
  du_k\,\bigg]\,\Pi'_k dh_k\;.  \e
We are permitted to make this small change of introducing $F(M)$
because our normal integrand for the second characteristic is
``protected'' from diverging for very small values of $\{\p_k\}$ by
the factor
    \b 1-\cos(\S_k\s h_k\s\p_k\s a^s)    \e
    in the numerator.
However, -- and unlike $\rho(\p,a)$ -- the weight function
$\rho_{1/M}(\p,a)$ is {\it integrable},
  and we now adjust $F(M)$ so that the
 integral is $M$ itself; i.e., we now assume that $F(M)$ is chosen so
 that
    \b \int\rho_{1/M}(\p,a)\,\Pi'_k\s d\p_k= M\;. \e
     As $M\ra\infty$, it is clear that $F(M)\ra0$; however, any
     further properties
     of the function $F(M)$ are not of particular interest
     to us.
With the normalization of $\rho_{1/M}(\p,a)$ fixed, we may recast
the expression for $\Psi_{1/M}^2(\p)$ into the form
   \b && \hskip-.7cm\Psi_{1/M}(\p)^2\equiv\bigg(\frac{a^s}{2\pi}\bigg)^{N'}\int\cos(\S_k\p_kh_k\s
  a^s)\\
  &&\hskip0cm\times\bigg[\,(1/M)\s\int\,\cos(\S_kh_k\s u_k \s a^s)\,\rho_{1/M}(u,a)\,\Pi'_k\s
  du_k\,\bigg]\,\Pi'_k dh_k\;,  \e
an equation which permits us to identify
  \b \Psi_{1/M}(\p)^2=\frac{1}{M}\,\frac{R(a)\s\exp[\s-\s U(\p,a)\s]}{\Pi'_k[\S_lJ_{k-l}\s
    \p_l^2\s+\s F(M)]^{1/2}}\;,  \e
    correct to order $O(1/M^2)$.

    This latter equation makes clear, {\it for large $M$}, that the
    ground-state distribution for this special case is just the
    weight function $\rho(\p,a)$ with its singularity regularized
    and then
    rescaled to have integral one. Apart from the
    denominator factor, the ground-state distribution $\Psi_{1/M}(\p)^2$ is proportional to
    $\exp[-U(\p,a)]$, which for the particular case in which
    $U(\p,a)$ is quadratic in the fields is highly suggestive of a
    free model. Here is further evidence supporting our designation
    of weight functions of the form
       \b \rho(\p,a)=\frac{R(a)\s\exp[\s-\s\S_{k,l}\s \p_k\s
   A_{k,l}\s \p_l\,a^{2s}\s]}{\Pi'_k[\S_lJ_{k-l}\s\p_l^2\s]^{1/2}}   \e
    as candidates for pseudofree models. ({\bf Remark:} It is noteworthy that certain
    highly specialized, soluble, nonrenormalizable models (see Chapters 9 \& 10 in
    \cite{book}) also exhibit ground-state distributions of a
    qualitatively similar form, including
    certain denominators not unlike our featured expression $\Pi'_k[\S_lJ_{k-l}\s
    \p_l^2\s+\s F(M)]^{1/2}$. In those cases, the analog of our function $U(\p,a)$
    was necessarily quadratic for the associated pseudofree models.)

    While the characteristic function of interest is easily regained
    as the $M{\rm th}$ power of the characteristic function $C(h)^{1/M}$,
     it is not nearly so easy to regain the ground-state distribution of
     interest since $\Psi(\p)^2$ is given by the {\it $M$-fold
    {\bf convolution} of the distribution $\Psi_{1/M}(\p)^2$}, followed, ultimately, by the
    limit $M\ra\infty$ to eliminate the error made in expanding the exponent.
    If we let the symbol $\s*\s$ denote convolution, then the
    computation just outlined is given by
      \b
      \Psi(\p)^2=\lim_{M\ra\infty}\,\Psi_{1/M}(\p)^2*
      \Psi_{1/M}(\p)^2*\cdots*\Psi_{1/M}(\p)^2 \;,  \e
      where the multiple convolution involves $M$ factors.

       After multiple
      convolutions, what
      happens to the original form of the expression  for
      the ground-state distribution $\Psi_{1/M}(\p)^2$
      is hard to predict. If $U(\p,a)$ is chosen as a quadratic function,
      which we have advocated for our pseudofree models, and if we momentarily assumed that
      the denominator in $\rho(\p,a)$ were absent, then the original distribution is a pure Gaussian, and
      under repeated convolutions it would still remain Gaussian; this is the usual situation for
      a strictly free model. However, the denominator of $\rho(\p,a)$ is exactly what
      separates the Gaussian and the generalized Poisson distributions, and
      therefore the denominator factor is
      of the utmost importance.

\subsection{A glimpse at the potential} We are unable to offer an
analytic form for the ground state $\Psi(\p)$, but we are able to
offer an analytic form for the ground state $\Psi_{1/M}(\p)$ valid
for large $M$. Of course, the latter ground state is {\it not} the
one of ultimate interest, but for someone starved for a look at any
form of a ground state, it may offer a glimpse into the problem that
is unavailable elsewhere. The potential associated with the ground
state $\Psi_{1/M}(\p)$ is given by the general formula, namely
   \b {\cal V}_{1/M}(\p)\equiv\frac{1}{2}\,
  {\sum}'_k\;\frac{a^{-s}}{\Psi_{1/M}(\p)}\,\frac{\d^2\Psi_{1/M}(\p)}{\d\p_k^2}\;, \e
which leads to the expression \b &&{\cal
  V}_{1/M}(\p)=\frac{1}{8}\s\s
a^{-s}\s{\sum_{k,s,t}}'\s\frac{J_{s-k}\s
  J_{t-k}\s \p_k^2}{[\S_m\s J_{s-m}\s\p^2_m+F(M)]\s[\S_n\s
  J_{t-n}\s\p_n^2+F(M)]} \\
  &&\hskip2.5cm-\frac{1}{4}\s\s a^{-s}\s{\sum_{k,t}}'\s\frac{J_{t-k}}{[\S_m\s
  J_{t-m}\s\p^2_m+F(M)]} \\
  &&\hskip2.5cm+\frac{1}{2}\s\s a^{-s}\s{\sum_{k,t}}'\s\frac{J_{t-k}^2\s\p_k^2}{[\S_m\s
  J_{t-m}\s\p^2_m+F(M)]^2}\s+\s{\tilde{\cal V}_{1/M}}(\p)\;, \e
  where ${\tilde{\cal V}_{1/M}}(\p)$ follows from terms which include
  either one or two derivatives of the unspecified term $U(\p,a)$. If we choose
     \b U(\p,a)=\S_{k,l}\s \p_k\s
   A_{k,l}\s \p_l\,a^{2s}\;,  \e
   it follows that the remainder potential ${\tilde{\cal V}_{1/M}}(\p)$ contains a quadratic
   potential term plus a term that arises from one derivative of $U$
   and one of the denominator. Observe well, that when factors of
   $\hbar$ are taken into account, the quadratic term is
   $O(\hbar^0)$, the cross term is $O(\hbar^1)$, while the term
   displayed in ${\cal V}_{1/M}(\p)$ above -- coming from two
   derivatives of the denominator factor --
    is $O(\hbar^2)$. In the
   classical limit, therefore, the only term that survives is the
   quadratic term; the other two terms are quantum corrections to
   the potential that arise strictly from the denominator factor
   present in the weight function $\rho_{1/M}(\p,a)$, and they
   make no contribution in the classical limit. Since the
   denominator carries no dimensional parameters, there is no
   possible way for it to contribute to the classical limit!

   Of course, we must not forget that we are only looking at the
   potential associated with $\Psi_{1/M}(\p)$ and not that associated with
   $\Psi(\p)$, the one of real interest. There is no simple way to tell how
   the potential ${\cal V}(\p)$ will appear after the multiple
   convolutions that $\Psi_{1/M}(\p)^2$ must undergo to become
   $\Psi(\p)^2$; nonetheless, there is no denying the feeling that
   something like the contribution displayed in the expression for
   ${\cal V}_{1/M}(\p)$ will survive, if for no other reason,
   than on dimensional grounds alone. However, that is only a guess.

    Understanding repeated
      convolutions of the distribution $\Psi_{1/M}(\p)^2$ for large $M$
      is one key to getting a handle on the ground-state
      distribution $\Psi(\p)^2$ of ultimate interest. Nevertheless, there may be another way
      to gain some information about $\Psi(\p)^2$, and that is to proceed {\it numerically}.

\subsection{Possible numerical studies}
 Whatever expression one chooses for the characteristic function $C(h)$ of the
  ground-state distribution $\Psi(\p)^2$, it follows that the
ground-state distribution itself is given simply by the inverse
Fourier transform, i.e.,
  \b  \Psi(\p)^2\equiv\bigg(\frac{a^s}{2\pi}\bigg)^{N'}\,\int\cos(\S_k\p_kh_k\s
  a^s)\,C(h)\,\Pi'_k\,d h_k\;. \e
If this expression could be numerically computed, at least for a
modest lattice size, we could then numerically determine
  $\Psi(\p)$, the potential ${\cal V}(\p)$, the lattice
  Hamiltonian $\H$, and finally the lattice action with which to
  perform
  Monte Carlo calculations for the various correlation functions of
  interest.

Admittedly, all that appears to be a rather formidable task.
  Therefore, one of the aims of this paper is to make this task appear more
  attractive.

\section{A Linear Superposition of Poisson \\Distributions}
Up to this point, we have placed all our bets on studying
generalized Poisson distributions as candidates for the ground-state
distribution. We now broaden the base of our inquiry to include a
much larger class of ground-state distributions based on linear
superpositions of the very class of distributions we have featured
so far.

In one-dimensional probability theory, it is noteworthy that a
linear superposition of even Gaussian distributions over their
variance parameter is dense in all possible even probability
distributions. This is likewise true for their characteristic
functions; namely, one may study the set of even characteristic
functions $C(-s)=C(s)$ that may be obtained from the relation
  \b C(s)=\int_0^\infty f(\tau)\,e^{-\tau\s s^2}\,d\tau  \e
  as $f$ ranges over (signed) generalized functions. It is easy to see that
  {\it all} even characteristic functions and thereby {\it all} even
  probability
  distributions may be obtained this way.\footnote{For example, let $f(\tau)$
  equal $\delta^{(m)}(\tau-1)$ to generate the elements $s^{2m}\s
  e^{-s^2}$, $m=0,1,2...$,
  take linear sums to construct the even order Hermite functions,
  and then note that sums of such functions are dense in all even functions, and
  hence in the subset of such functions that are characteristic functions.}
  A similar question can be raised about linear sums of even Poisson
  distributions, or equivalently about their even characteristic
  functions. In equation form, this latter question reduces to the class of
  characteristic functions that can be reached by linear superposition (say) of the
  limited set of Poisson characteristic functions
   \b C(s)=\int_0^\infty\int_0^\infty f(\tau,\omega)\,\exp\{-\tau\tint[1-\cos(s\s
   u)]\,e^{-\omega\s u^2}\,du/u^2\s\}\,d\tau\,d\omega \e
   again as $f$ ranges over (signed) generalized functions. Since we can
   reconstruct the Gaussian case from the Poisson case, it is clear
   that once again such linear combinations are dense in the set of
   {\it all} even characteristic functions and thereby {\it all}
   even probability distributions.

 In this section we study
expressions for the lattice-space characteristic function associated
with the (even) ground-state distribution in the form of a linear
superposition of generalized Poisson characteristic functions given
by
 \b&& C_{(K)}(h)\equiv\int \cos(\Sigma'_k\s h_k\s\p_k\s a^s)\,\Psi_{(K)}(\p)^2\,
\Pi'_k\s d\p_k\\
   && \hskip1.5cm\equiv\int d\tau\,K(\tau)\,\exp\{-\tint[1-
\cos(\Sigma'_k\s h_k\s\p_k\s a^s)]\s\rho_\tau(\p,a)\,\Pi'_k\s
d\p_k\}\;.  \e The basic ingredient in this superposition is the
expression
  \b C_\tau(h)\equiv\exp\{-\tint[1-\cos(\Sigma'_k\s h_k\s\p_k\s a^s)]
  \s\rho_\tau(\p,a)\,\Pi'_k\s d\p_k\}\;,  \e
which, as discussed in Sec.~4, is the characteristic function for a
generalized Poisson distribution, an infinitely divisible
characteristic function.  Note well that we do not require that
$C_{(K)}(h)$ is an infinitely divisible distribution itself;
instead, we use a suitable set of infinitely divisible distributions
as ``building blocks'' for $C_{(K)}(h)$.

Since a limit of Poisson distributions exists that converge to a
Gaussian distribution, we may allow for such a possibility in the
continuum limit as $a\ra0$. In this sense, the expression given for
our basic element incorporates both Poisson and Gaussian
distributions, and therefore potentially includes a wide class of
infinitely divisible distributions. With regard to the superposition
by an integral over $\tau\in{\mathbb R}^p$, for some $p$, one may
focus initially on positive measures $K(\tau)\s d\tau$ (thus
including $\delta$-functions) such that $\tint K(\tau)\s d\tau=1$.
However, it is also possible to extend such examples to suitable
signed measures $K(\tau)\s d\tau$ such that $\tint |K(\tau)|\s
d\tau<\infty$, $\tint K(\tau)\s d\tau=1$, and $\Psi(\p)^2>0$; we
content ourselves with a dense set of results and do not pursue
examples where $K(\tau)$ is a generalized function. Our purpose in
this exercise is to increase, broadly, we hope, the scope of
characteristic functions and thereby of ground state distributions
that may be considered. We make no claim that our limited family of
superpositions is exhaustive of all such distributions (unlike the
one-dimensional case).

By assumption, the continuum limit of our lattice-space formulation
$C_{(K)}(h)$ is intended to retain the basic structure of the
separate Poisson characteristic functions that make up the linear
combination. This implies that in the continuum limit the resultant
expression for $E_{(K)}(h)$ is generally non-Gaussian. It is our
proposal to look for the ground-state distribution of various scalar
models within the family of characteristic functions offered above,
and in particular for those models that are pseudofree models (thus
having a limited number of parameters). As emphasized previously,
such pseudofree models may serve as starting points for a
perturbation analysis for more interesting interacting models.

On one hand, it may even be possible to find the ground-state
distribution for some interacting models within the family obtained
by linear superposition. On the other hand, although the class of
ground-state distributions $\Psi(\p)^2$ that may be described by the
linear superposition of infinitely divisible distributions may be
relatively large, the reader may well ask by what reason could we
expect to find an acceptable description of (say) a $\varphi^4_n$,
$n\ge5$, model within that particular class of characteristic
functions.

 To address this question consider the following one-dimensional quantum mechanical problem:
 A system with classical Hamiltonian $H(p,q)$ is quantized by adopting the quantum Hamiltonian
  \b \H=H(P,Q)+\hbar\s Y(P,Q)\;, \e
where $Y$ is usually present to account for possible factor ordering
ambiguity. However, $Y$ could in fact be a quite general operator
(so long as $\H$ is self adjoint) because in the classical limit in
which $\hbar\ra0$ the original classical Hamiltonian $H(p,q)$ is
recovered.  This kind of ambiguity is always present in quantum
theory, and the traditional way to deal with it is to ``appeal to
experiment''. ({\bf Remark:} For example, such additional terms are
surely present in any analysis of nonrenormalizable models based on
the counterterms suggested by perturbation theory.)

It is possible that approaching a wide class of scalar models
through the family of characteristic functions represented by
$C_{(K)}(h)$ may correspond to a realization of certain models with
the presence of selected counterterms (the analogue of $Y(P,Q)$
above). Since a principal goal of the present study is to achieve
nontriviality, any satisfactory set of counterterms will suffice
initially; limitations that may arise from an appeal to experiment
can be considered later.

In the following section we will discuss correlation functions for
the case of a ground-state distribution which is a generalized
Poisson distribution. That same discussion can easily be extended to
also include those distributions that are given by the linear
superposition of Poisson distributions.

  \section{Correlation Functions}
  Even without fully defining our
  choice for a full-time, lattice-space action function including
  the required auxiliary potential, we can nevertheless draw some
  important general conclusions. In particular, let us show that the
  full-time correlation functions can be
  controlled by their sharp-time behavior along with a suitable
  choice of test sequences.
\subsection{Correlation function bounds for general \\distributions}
   Let the notation
    \b  \p_u\equiv \Sigma_k u_k\s\p_k\,a^n \e
denote the full-time summation over all lattice fields where
$\{u_k\}$ denotes a suitable test sequence. We also separate out the
temporal part of this sum in the manner
   \b \p_u\equiv\Sigma_{k_0}a \,\p_{u'}\equiv
\Sigma_{k_0}\s a\,\S_k u_k\s\p_k\,a^s\;. \e Observe that the
notation $\p_{u'}$ (with the prime) implies a summation over {\it
only} the spacial lattice points for a fixed (and implicit) value of
the temporal lattice value, $k_0$.

Let the notation $\<(\s\cdot\s)\>$ denote full-time averages with
respect to the field distribution determined by the lattice action,
and then let us consider full-time correlation functions such as
 \b  \hskip.2cm\<\s\p_{u^{(1)}}\s\p_{u^{(2)}}\s\cdots\s\p_{u^{(2q)}}\s\>
 = \Sigma_{k_0^{(1)},k_0^{(2)},\s\dots\s,k_0^{(2q)} }\,a^{2q}\,\<\s\p_{u'^{(1)}}
\s\p_{u'^{(2)}}\s\cdots\s\p_{u'^{(2q)}}\s\> \;, \hskip.4cm q\ge1\;,
\e where the expectation on the right-hand side is over products of
fixed-time summed fields, $\p_{u'}$, for possibly different times,
which are then summed over their separate times.
 All odd correlation functions are assumed
    to vanish, and furthermore, $\<\s1\s\>=1$ in this normalized
    spacetime lattice field distribution.
It is also clear that
  \b  |\<\s\p_{u^{(1)}}\s\p_{u^{(2)}}\s\cdots\s\p_{u^{(2q)}}\s\>|
 \le \Sigma_{k_0^{(1)},k_0^{(2)},\s\dots\s,k_0^{(2q)} }\,a^{2q}\,|\<\s\p_{u'^{(1)}}
\s\p_{u'^{(2)}}\s\cdots\s\p_{u'^{(2q)}}\s\>| \;.  \e At this point
we turn our attention toward the spatial sums alone.

    We appeal to straightforward inequalities of the general
    form
      \b \s\<A\s B\>^{2}\le \<\s A^2\s\>\s\<\s B^2\s\>\;.  \e
   In particular, it follows that
    \b
    \<\s\p_{u'^{(1)}}\s\p_{u'^{(2)}}\s\p_{u'^{(3)}}\s\p_{u'^{(4)}}\>^2
     \le\<\s\p^2_{u'^{(1)}}\s\p^2_{u'^{(2)}}\>\s\<\s\p^2_{u'^{(3)}}\s\p^2_{u'^{(4)}}\>\;,
     \e
     and, in turn, that
      \b \<\s\p_{u'^{(1)}}\s\p_{u'^{(2)}}\s\p_{u'^{(3)}}\s\p_{u'^{(4)}}\>^4
      \le\<\s\p^2_{u'^{(1)}}\s\p^2_{u'^{(2)}}\>^2\s\<\s\p^2_{u'^{(3)}}\s\p^2_{u'^{(4)}}\>^2
      \le\<\s\p^4_{u'^{(1)}}\>\s\<\p^4_{u'^{(2)}}\>\s\<\s\p^4_{u'^{(3)}}\>\s\<\p^4_{u'^{(4)}}\>\;.
      \e
      By a similar argument, it follows that
      \b
      |\<\s\p_{u'^{(1)}}\s\p_{u'^{(2)}}\s\cdots\s\p_{u'^{(2q)}}\s\>|
      \le \Pi_{j=1}^{2q}\,[\s\<\p^{2q}_{u'^{(j)}}\>]^{1/2q}\;, \e
      which has bounded any particular mixture of spatial correlation functions at
      possibly different times,
      by a suitable product of higher-power expectations each of
      which involves field values ranging over a spatial level, all at a single
      fixed lattice time.
     By time translation invariance of the various single time correlation functions
      we can assert that
        \b \<\s \p^{2r}_{u'^{(j)}}\s\>\;, \e
        which is defined at time $k_0^{(j)}$, is actually independent of
        the time and, therefore, the result could be calculated at any
        fixed time. In particular, we can express such correlation functions as
        \b  \<\s\p^{2q}_{u'}\s\>=\int
        \p^{2q}_{u'}\,\Psi(\p)^2\,\Pi'_k\,d\p_k \;.\e
        Adding this relation to those established above completes a bound on the multi-time correlation
        function in terms of moments of the ground-state distribution.
\subsection{Correlation function bounds for Poisson \\distributions}
Observe that the bounds we have discussed up to this point apply to
arbitrary ground-state distributions $\Psi(\p)^2$. Let us now
specialize to generalized Poisson distributions with which we shall
be able to make additional and more specific remarks.

For Poisson distributions, we note that the moments of the
ground-state distribution for the correct (but unknown) ground-state
may be directly related to the {\it truncated} moments for the
ground-state distribution, which in turn are directly determined by
the ``moments'' of the weight function $\rho(\p,a)$. Let us recall
the notation introduced earlier, namely
 \b (\p^{2p}_{u'})\equiv \tint \p^{2p}_{u'}\,\rho(\p,a)\,\Pi'_k\s d\p_k\;,
 \hskip1cm p\ge1\;,\e
 which are not moments but noments (since $\tint \rho(\p,a)\,\Pi'_k\s
 d\p_k=\infty$). These noment expressions coincide exactly with the
 truncated moments of the ground state distribution. Recall that
 ordinary and truncated moments are related by the generating
 function relation
  \b \<\s e^{\a\s\p_{u'}}\s\>=\exp[\s{\<\s e^{\a\s\p_{u'}}-1\s\>^T}\s]\;, \e
which holds for all $\a$, where $T$ denotes truncated. Therefore, in
our case,
   \b \<\s \p^{2p}_{u'}\s\>^T\equiv (\s\p^{2p}_{u'}\s)\;,\hskip1cm p\ge1\;. \e
For example, we list a few relations that follow from this
connection:
  \b &&\<\s \p^{2}_{u'}\s\>=(\s\p^{2}_{u'}\s)\;,\\
  &&\<\s \p^{4}_{u'}\s\>=(\s\p^{4}_{u'}\s)+3\s(\s\p^{2}_{u'}\s)^2\;,\\
  &&\<\s\p^{6}_{u'}\s\>=(\s\p^{6}_{u'}\s)+15\s(\s\p^{2}_{u'}\s)(\s\p^{4}_{u'}\s)+15\s(\s\p^{2}_{u'}\s)^3\;,\\
  &&\<\s\p^{8}_{u'}\s\>=(\s\p^{8}_{u'}\s)+28\s(\s\p^{2}_{u'}\s)(\s\p^{6}_{u'}\s)+35\s(\s\p^{4}_{u'}\s)^2\\
  &&\hskip2.28cm +\s\s210\s(\s\p^{2}_{u'}\s)^2(\s\p^{4}_{u'}\s)+105\s(\s\p^{2}_{u'}\s)^4\;,\e
etc.

The conclusion of this exercise is that all of the multi-time
correlation functions are bounded by terms that are composed from
the noments determined from the weight function $\rho(\p,a)$. If we
can choose $\rho(\p,a)$ so that it has a suitable continuum limit,
and in  such a way that the noments remain finite in that limit,
then we will have obtained a bound on every full-time correlation
function in the continuum limit. Convergence (of subsequences, if
necessary) of these correlation functions follows. Moreover, the
ground-state distribution stays firmly within the family of Poisson
distributions and, by working with the various noments
$(\s\p^{2p}_{u'}\s)$, we have avoided the fate of a Gaussian, i.e.,
trivial, limiting distribution.

\section{From Noments to Genuine Averages}
\subsection{A two-point normalization}
The discussion regarding correlation functions has shown that many
functions of interest can be evaluated exactly, or are suitably
bounded, by terms involving various noments such as
  \b (\p^{2q}_{u'})=R(a)\int (\S_k u_k\s\p_k\s a^s)^{2q}\,\frac{e^{-\S_{k,l}\p_k\s
  A_{k,l}\s
  \p_la^{2s}}}{\Pi'_k[\S_lJ_{k-l}\p_l^2]^{1/2}}\,\Pi'_kd\p_k\;, \e
  where $q\ge1$ and  $u_k$ denotes a test sequence here used in space alone.
  Expansion of the power $2q$ leads to expressions of the sort
  \b  (\s\p_{l_1}\s\p_{l_2}\ldots\p_{l_{2q}}\s)=R(a)
  \int\s\p_{l_1}\s\p_{l_2}\ldots\p_{l_{2q}}\s\,\frac{e^{-\S_{k,l}\p_k\s
  A_{k,l}\s
  \p_la^{2s}}}{\Pi'_k[\S_lJ_{k-l}\p_l^2]^{1/2}}\,\Pi'_kd\p_k\;.  \e
  It is these noments that we wish to discuss in this section with
  the possible aim of
approximately evaluating them using Monte Carlo techniques.

In order to invoke Monte Carlo calculational methods, however, it is
clear that we need to introduce a normalized probability measure,
and for this purpose it will suffice to fix the factor $R(a)$, at
least implicitly. We are free to fix this normalization based on the
freedom involved in the usual field-strength renormalization. And we
do so by selecting a carefully chosen and suitably weighted
two-point function and then declare that the noment of this
expression is unity.

To this end, let us choose $Y_{k,l}$ -- where, in this case, $k$ and
$l$ involve only the {\it spatial} components -- as the elements of
a symmetric matrix $Y\equiv \{Y_{k,l}\}$ which is positive definite.
Furthermore the elements $Y_{k,l}$ are chosen so that their ``high
frequency'' components are suitably reduced. We will offer two
examples of what we deem to be suitable matrices, and from those it
should become clear just what we have in mind regarding the ``high
frequency'' components. The first example we choose for $Y_{k,l}$ is
suitable for a lattice formulation of the problem, such as may arise
in Monte Carlo calculations, where the lattice size $L$ and the
lattice spacing $a$ may vary, but remain finite, and no attempt at
approaching the continuum limit is contemplated; this example will
be used in Sec.~7.1. The second example we choose for $Y_{k.l}$ is
one designed to be use for an approach to the continuum limit; the
second example will be used in Sec~7.2 in which the continuum limit
is discussed.

Observe that the unit matrix $\delta_{k,l}$ is positive definite,
but it does not discriminate between ``low'' and ``high''
frequencies. To describe the unit matrix in other terms, we first
assume that $L$, the number of lattice points along each axis, is
odd, or in other words, $L=2P+1$, where $P$ is a positive integer.
We choose two integers $r$ and $s$, where $-P\le r\le P$, and
similarly for $s$, and observe that
\b\frac{1}{L}\sum_{m=-P}^P\,e^{2\pi
  i(r-s)\s m/L}=\delta_{r,s}\;.\e
  Now, as a preliminary to making our choice for $Y_{k,l}$, we dampen the weight of the
  high frequencies in the previous sum by defining
    \b Z_{r,s}\equiv \frac{1}{L}\sum_{m=-P}^P\,e^{-|m|\s T}\,e^{2\pi
  i(r-s)\s m/L}\;, \e
  where $T>0$ is a damping parameter to be chosen.
  Clearly, the resultant function is still positive definite as
  desired. Indeed, the sum can be evaluated in closed form as
    \b \frac{\{1-e^{-2T}-2\s e^{-(P+1)T}\s\cos[2\pi\s\Delta\s(P+1)/L]
    +2\s e^{-(P+2)T}\s\cos[2\pi\s\Delta\s P/L]\}}{L\s \{1-2\s e^{-T}\s\cos[2\pi\Delta/L]+e^{-2T}\}}\;,\e
where $\Delta\equiv r-s$. Finally, to define our first choice for
$Y_{k,l}$, we let
      \b Y_{k,l}\equiv\prod_{j=1}^s Z_{k_j,l_j}\;,  \e
where, as noted above, $k$ and $l$ involve only the spatial
components. If, on the other hand, $L$ is even, i.e., $L=2P$, for
integral $P>0$, then we choose
  \b  Z_{r,s}\equiv \frac{1}{L}\sum_{m=-P+1}^P\,e^{-|m|\s T}\,e^{2\pi
  i(r-s)\s m/L}\;, \e
which has the closed form
  \b Z_{r,s}=[1-e^{-PT}(-1)^\Delta]\bigg\{\frac{2\s[1-e^{-T}\s\cos(2\pi\Delta/L)]}{[1-2\s
  e^{-T}\cos(2\pi\Delta/L)+e^{-2T}]}-1\bigg\}\;. \e
  The quantity $Y_{k,l}$ is constructed from
  $Z_{r,s}$ in the same way as for the odd $L$ case.

 Although not needed immediately, we also choose our second
form for $Y_{k,l}$ suitable for discussing the continuum limit. For
the following, let $z\in{\mathbb R}$, and let the orthonormal set of
Hermite functions be called $h_n(z)$, $0\le n<\infty$, with the
usual property that
 \b  \sum_{n=0}^\infty h_n(z'')\, h_n(z')=\delta(z''-z')\;. \e
 We dampen the high frequencies once again and define
 \b  Z(z'',z')\equiv\sum_{n=0}^\infty\,e^{-n\s T}\, h_n(z'')\, h_n(z')\;, \e
 where again $T>0$ is a parameter to be chosen. This sum can be evaluated in closed form as
   \b  Z(z'',z')=\frac{1}{\sqrt{2\pi T}}\,\exp[-(1/2)(z''^2+z'^2)\coth(T)+z''\s z'\s {\rm csch}(T)]\;.\e
Clearly, the kernel $Z(z'',z')$ is positive definite. Finally, we
define the function $Y(x,y)$ -- where both
$x=\{x_1,x_2,\ldots,x_s\}$ and $y=\{y_1,y_2,\ldots,y_s\}$ are in
${\mathbb R}^s$ -- by the expression
  \b Y(x,y)\equiv \prod_{j=1}^s Z(x_j,y_j) \;.\e
  Note that since each Hermite function qualifies as a test
  function, we may refer to $Z(z'',z')$ and $Y(x,y)$ as positive
  definite ``test kernels''.
  The given choice $Y(x,y)$ is suitable for the continuum limit for the
  whole space ${\mathbb R}^s$, but if one chooses to use it for a large
  lattice, before taking the continuum limit, it would be acceptable to use
  \b  Y_{k,l}\equiv Y(ka,la)\;, \e
  namely, the values that $Y(x,y)$ assumes on the lattice points
  themselves.

Having chosen a suitable, positive definite matrix $Y_{k,l}$, as
discussed above, we are in a position to establish our normalization
criterion. In particular, for a given, fixed, choice of the matrix
$Y$, we (arbitrarily) declare that the noment
     \b R(a)\int \S_{k,l}\s\p_k\s Y_{k,l}\s\p_l\s a^{2s}\,\frac{e^{-\S_{k,l}\p_k\s
  A_{k,l}\s
  \p_la^{2s}}}{\Pi'_k[\S_lJ_{k-l}\p_l^2]^{1/2}}\,\Pi'_kd\p_k=1\;. \e
 This
  normalization condition fixes the constant $R(a)$. In effect, we
  are saying that a certain two-point function with a suitably
  chosen positive definite weighting has a noment of unity. We can put this normalization
  to good use for us as follows.

  For all $q\ge1$, we rewrite the noment of general interest as
   \b &&(\s\p_{l_1}\s\p_{l_2}\ldots\p_{l_{2q}}\s)=\frac{R(a)
  \int\s\p_{l_1}\s\p_{l_2}\ldots\p_{l_{2q}}\s\,\frac{e^{-\S_{k,l}\p_k\s
  A_{k,l}\s
  \p_la^{2s}}}{\Pi'_k[\S_lJ_{k-l}\p_l^2]^{1/2}}\,\Pi'_kd\p_k}{R(a)\s
  \int (\S_{k,l}\s\p_k\s Y_{k,l}\s\p_l\s a^{2s})\,\frac{e^{-\S_{k,l}\p_k\s
  A_{k,l}\s
  \p_la^{2s}}}{\Pi'_k[\S_lJ_{k-l}\p_l^2]^{1/2}}\,\Pi'_kd\p_k}\\
  && \hskip1cm=\frac{
  \int\s\frac{\p_{l_1}\s\p_{l_2}\ldots\p_{l_{2q}}}{\S_{k,l}\s\p_k\s Y_{k,l}\s\p_l\s a^{2s}}\s\,
  \frac{\S_{k,l}\s\p_k\s Y_{k,l}\s\p_l\s a^{2s}\s e^{-\S_{k,l}\p_k\s
  A_{k,l}\s
  \p_la^{2s}}}{\Pi'_k[\S_lJ_{k-l}\p_l^2]^{1/2}}\,\Pi'_kd\p_k}{
  \int \frac{\S_{k,l}\s\p_k\s Y_{k,l}\s\p_l\s a^{2s}\s e^{-\S_{k,l}\p_k\s
  A_{k,l}\s
  \p_la^{2s}}}{\Pi'_k[\S_lJ_{k-l}\p_l^2]^{1/2}}\,\Pi'_kd\p_k}\\
  &&\hskip1cm\equiv\int\s\frac{\p_{l_1}\s\p_{l_2}\ldots\p_{l_{2q}}}
  {\S_{k,l}\s\p_k\s Y_{k,l}\s\p_l\s a^{2s}}\s\,d\sigma(\p)\;.\e
This last equation establishes the noment of interest as a {\it
genuine average} involving a normalized probability measure
 \b &&d\sigma(\p)\equiv
 \frac{[\s\S_{k,l}\s\p_k\s Y_{k,l}\s\p_l\s a^{2s}\s]\s e^{-\S_{k,l}\p_k\s
  A_{k,l}\s
  \p_la^{2s}}}{\Pi'_k[\S_lJ_{k-l}\p_l^2]^{1/2}}\,\Pi'_kd\p_k\\
  &&\hskip1.4cm\times\s{\bigg/}
  \int\frac{[\s\S_{k,l}\s\p_k\s Y_{k,l}\s\p_l\s a^{2s}\s]\s e^{-\S_{k,l}\p_k\s
  A_{k,l}\s
  \p_la^{2s}}}{\Pi'_k[\S_lJ_{k-l}\p_l^2]^{1/2}}\,\Pi'_kd\p_k \;. \e
It is the final expression for
$(\s\p_{l_1}\s\p_{l_2}\ldots\p_{l_{2q}}\s)$ we would like to
evaluate approximately by means of
  conventional Monte Carlo methods.

  We have arrived at the expression of noments as genuine averages on the basis of a convenient choice
  of normalization given above. Note that if instead we had chosen
  to normalize matters so that
  \b R(a)\s \int \S_{k,l}\s\p_k\s Y_{k,l}\s\p_l\s a^{2s}\,\frac{e^{-\S_{k,l}\p_k\s
  A_{k,l}\s
  \p_la^{2s}}}{\Pi'_k[\S_lJ_{k-l}\p_l^2]^{1/2}}\,\Pi'_kd\p_k=b\;, \e
  rather than $b=1$ as above, then the final result would have been
    \b (\s\p_{l_1}\s\p_{l_2}\ldots\p_{l_{2q}}\s)=b\s\int\s\frac{\p_{l_1}\s\p_{l_2}\ldots\p_{l_{2q}}}
    {\S_{k,l}\s\p_k\s Y_{k,l}\s\p_l\s a^{2s}}\s\,d\sigma(\p)\;,  \e
    where the normalized probability measure $d\sigma(\p)$ is
    unchanged. Hereafter, we assume that
    $b=1$.

    It should be noted that there is a special advantage derived from
    the fact that the noments can be expressed as suitable averages
    over a particular probability distribution. This advantage
    follows from the fact that inequalities such as
      \b  \int\s\bigg[\frac{(\p_{l_1}\s\p_{l_2})}
    {\S_{k,l}\s\p_k\s Y_{k,l}\s\p_l\s a^{2s}}\bigg]^2\s\,d\sigma(\p)\ge\bigg[\int\s\frac{(\p_{l_1}\s\p_{l_2})}
    {\S_{k,l}\s\p_k\s Y_{k,l}\s\p_l\s a^{2s}}\s\,d\sigma(\p)\bigg]^2  \e
hold.
\subsection{Continuum limit}
 Finally, we briefly take up the question regarding the continuum
limit. There are several important and distinct aspects of the
continuum limit, namely, not only $a\ra0$ but $N'\ra\infty$, and in
fact, it is also necessary that the spacial lattice volume $N'\s
a^s\ra\infty$. Let us first discuss the situation for large $N'$.

When dealing with a large number of integration variables, i.e.,
$N'\gg1$, it
   is important to note that there are important -- even profound -- differences between
   Gaussian-like integrals and Poisson-like integrals. In presenting
   the following discussion we closely follow a section of
   \cite{klau} dealing with an idealized set of examples. For all $p\ge1$, consider
   the two sets of integrals
     \b I_G(2p)=\int(\S_k\p^2_k)^p\s e^{-A\S_k\p_k^2}\,\Pi'_k d\p_k\;,\e
     as representative of the Gaussian-like expressions, and
      \b I_P(2p)=\int(\S_k\p_k^2)^p\s
      e^{-A\S_k\p_k^2}\s[\S_k\p_k^2]^{-N'/2}\,\Pi'_k d\p_k\;, \e
      as representative of the Poisson-like expressions.
      These integrals are ``caricatures'' of the ones
      we study, but the results are nevertheless informative.

      To help
      in our study, let us introduce {\it hyper-spherical
      coordinates} \cite{klau} (c.f., also \cite{iande}) defined by
        \b &&\p_k\equiv\k\eta_k\;,\hskip.5cm
        0\le\k<\infty\;,\hskip.5cm -1\le\eta_k\le1\;, \\
        &&\hskip1.4cm\S_k\eta_k^2\equiv1\;,\hskip.5cm \S_k\p_k^2\equiv\k^2\;. \e
        Here $\k$ denotes an overall radius variable while
        $\{\eta_k\}$ denotes an $N'$-dimensional direction field. In terms of these
        variables, the Gaussian-like integrals take on the form
          \b I_G(2p)=2\int\k^{2p}\s
          e^{-A\k^2}\s\k^{N'-1}\,d\k\,\delta(1-\S_k\eta_k^2)\,\Pi'_k d\eta_k\;.\e
          Evaluation of this expression for large $N'$ is dominated
          by the factor $\k^{N'-1}$, and a steepest descent method
          can be used to evaluate the integral over $\k$ to a
          suitable accuracy. To leading order, it follows that the stationary point is
          given by $\k=(N'/2A)^{1/2}$. As a consequence, for each
          value of $A$, the integrand is supported on a
          disjoint set of $\k$ as $N'\ra\infty$. This well-known fact leads to
          divergences in perturbation calculations. For example, let
          us calculate
            \b I^\star_G(2)=\int(\S_k\p_k^2)\s e^{-A^\star\S_k\p_k^2}\,\Pi'_k d\p_k\e
            for a different value of $A$ by the perturbation series,
            \b I^\star_G(2)=I_G(2)-\Delta A\,I_G(4)+\half(\Delta
            A)^2\,I_G(6)-\cdots\;, \e
            where $\Delta A\equiv A^\star-A$. Since
            $I_G(2p)/I_G(2)\propto N'^{(p-1)}$, this series exhibits
            {\it divergences} as $N'\ra\infty$.

            Let us now consider the Poisson-like integrals expressed
            in the same hyper-spherical coordinates. It follows that
              \b  I_P(2p)=2\int\k^{2p}\s
              e^{-A\k^2}\s[\k^2]^{-N'/2}\s\k^{N'-1}\,d\k\,
              \delta(1-\S\eta_k^2)\,\Pi'd\eta_k\;.\e
Here we see no large $\k$-power in the integrand, and it follows,
e.g., that
    \b \frac{I_P(4)}{I_P(2)}=\frac{\tint\k^3\s e^{-A\k^2}\,d\k}{\tint \k\s
    e^{-A\k^2}\,d\k}=\frac{1}{A}\;, \e
which is finite and independent of $N'$. Hence a perturbation
calculation of the change of $A$ to $A^\star$ for $I_P(2)$, for
example, involves no divergences as $N'\ra\infty$! Further examples
of the difference of Gaussian- and Poisson-like integrals is offered
in \cite {klau}, but the main point has already been made showing
the profound difference between these two types of integral for
large $N'$.

How do these examples impact on our main discussion? Consider the
expression for a simple noment given by
  \b  (\p_t^2)=R(a)\s\int \p_t^2\,\frac{e^{-\S_{k,l}\s\p_k\s
  A_{k,l}\s\p_l\s a^{2s}}}{\Pi'_k[\S_l\s J_{k-l}\s\p_l^2]^{1/2}}\;\Pi'_k\s d\p_k\;.
  \e
Expressed in hyper-spherical coordinates, this integral becomes
  \b  (\p_t^2)=2\s R(a)\s\int \k^2\eta_t^2\,\frac{e^{-\k^2\S_{k,l}\s\eta_k\s
  A_{k,l}\s\eta_l\s a^{2s}}}{\Pi'_k[\S_l\s J_{k-l}\s\eta_l^2]^{1/2}}\,\k^{-1}
  \,d\k\,\delta(1-\S_k\eta_k^2)\,\Pi'_k\s
  d\eta_k\;,\e
  and we see an analogous absence of the parameter $N'$ regarding
  the integral over $\k$. This favorable situation has arisen from
  the presence of a denominator factor which, although different in
  detail, has an overall power of $\k$ identical to that of the
  elementary examples for $I_P(2)$. Ideally, one might like to
  choose a denominator factor that is as local as possible on the
  lattice, and that would mean choosing a set of $\{\beta_k\}$
  parameters of the form $\beta_k=\beta\,\delta_{k,0}$. This choice
  has the desired locality property, but it does not lead to
  an integrable denominator, as noted previously and which is further
  discussed in the Appendix. The choice of
  $\beta_k=J_k$ -- that is, our featured $\kk$ set of values -- is a suitable compromise
  that has locality in the continuum limit but retains enough
  lattice spread to always ensure an integrable denominator. In calculations
  of our principal interest, and  with our featured choice of the denominator factor,
  we expect to benefit from the same
  lack of divergences due to the missing large $N'$ dependence of
  the hyper-spherical radius $\k$ as was exhibited in the elementary
  examples.

Next we consider the problems associated with the limit $a\ra0$
along with $N'\ra\infty$, ideally, so that $N'a^s\ra\infty$, but at
the very least, so that $N'a^s\ra V'$, where $0<V'<\infty$, and
preferably $V'$ is ``large'' in some sense. To appreciate the kind
of problems faced it is useful to examine first the usual situation
for a ``free'' theory and its continuum limit.

Consider the lattice formulation for the ground-state distribution
of the free ($=$ Gaussian) theory given by
  \b C_F(h)=M\int e^{i\S_k h_k\p_k\s a^s}\,e^{-\S_{k,l}\p_k
  B_{k,l}\p_l\s a^{s}}\,\Pi'_k d\p_k=e^{-(1/4) \S_{k,l}h_kB^{-1}_{k,l} h_l\s a^s}\;,  \e
  where $M$ denotes a normalization so that $C_F(0)=1$. It is clear
  in this case that the two-point function is given by
     \b  \<\p_r\s \p_s\>=M\int \p_r\s \p_s\,e^{-\S_{k,l}\p_k
  B_{k,l}\p_l\s a^{s}}\,\Pi'_k d\p_k=\half B^{-1}_{r,s}\,a^{-s}\;, \e
  i.e., essentially the $r,s$ matrix element of $B^{-1}$, the inverse matrix
  to $B$. A suitable continuum limit of this expression starts by
  considering the smeared expression
    \b \<\s(\S_r\s u_r\s\p_r\s a^s)^2\s\>=\half\S_{r,s}u_r\s B^{-1}_{r,s}\s u^s\,a^{s}\;, \e
  for sequences $\{u_r\}$, for example, such that $\S_r u_r^2\s a^s=1$.
  In this case, the continuum limit should assume the form
    \b \<\s(\tint u(x)\s\p(x)\,d^s\!x)^2\s\>=\half\tint u(x)\s u(y)
    F(x,y)\,d^s\!x\s d^s\!y\;,  \e
    for some {\it generalized function} $F(x,y)$ and for all
    functions
    $u(x)\in C^\infty_0$, for example, for which $\tint
    u(x)^2\,d^s\!x=1$. The key concept is the limiting behavior of
    the lattice two-point function to a suitable generalized
    function in the continuum limit, a well-known property for a
    free theory. Observe that the properties of $F(x,y)$ are
    directly determined by the properties of the matrix $B^{-1}$, the inverse
    of the  matrix $B$.

Now let us consider a similar limit for the ground-state
distribution determined by a generalized Poisson distribution. In
particular, we first focus on the two-point function
   \b P_{r,s}\equiv\<\p_r\s \p_s\>=R(a)\int \p_r\s\p_s\,\frac{e^{-\S_{k,l}\p_k\s A_{k,l}\s
   \p_l s^{2s}}}{\Pi'_k[\S_lJ_{k-l}\s\p_l^2]^{1/2}}\,\Pi'_k\s d\p_k\;,  \e
   where, as discussed above, normalization is set by
     \b R(a)^{-1}=\int [\S_{k,l}\s\p_k\s Y_{k,l}\s\p_l\s a^{2s}]\,\frac{e^{-\S_{k,l}\p_k\s A_{k,l}\s
   \p_l s^{2s}}}{\Pi'_k[\S_lJ_{k-l}\s\p_l^2]^{1/2}}\,\Pi'_k\s d\p_k\;. \e
   As in the Gaussian example, we now consider the expression
   \b  \<(\S_ru_r\s\p_r\s a^s)^2\>=\S_{r,s}\s u_r P_{r,s}\s u_s
   a^{2s}\;,  \e
   and, for a satisfactory continuum limit, we ask that as $a\ra0$
   and $N'a^s\ra\infty$, or at least $N'a^s\ra V'\gg1$, we find that
     \b \<\s(\tint u(x)\s\p(x)\,d^s\!x)^2\s\>=\tint u(x)\s u(y)\s
     G(x,y)\,d^s\!x\s d^s\!y\;, \e
     for a {\it suitable generalized function} $G(x,y)$, say, for the same
     set of smooth functions $u$ considered above. In particular, according to
     our normalization procedure for determining $R(a)$, we have required that the
     continuum choice of the positive definite test kernel, $Y(x,y)$, such as the
     example introduced in Sec.~7.1, satisfy
   \b \tint \,Y(x,y)\s
     G(x,y)\,d^s\!x\s d^s\!y=1\;.  \e

     Note that, quite
     unlike the Gaussian case, the matrix $A$ is only {\it indirectly} involved
     in the determination of the generalized function $G(x,y)$.
     Indeed, to have a suitable form for $G(x,y)$, there is no
     guarantee that the matrix $A$ has any clear continuum limit by
     itself, i.e., there is no requirement that
       \b \S_{k,l} \p_k\s A_{k,l}\s\p_l\s a^{2s}\ra\tint \p(x)\s
       A(x,y)\s\p(y)\,d^s\!x\s d^s\!y  \e
       for a suitable generalized function $A(x,y)$ in the continuum
       limit. All that is required is that $G(x,y)$ be suitable --
       as well as suitable distributional behavior for
       all the higher-order correlation functions as well!

       It would be helpful to build up some
       computational experience for correlation functions for the
       generalized Poisson distributions in order to eventually
       help select matrices such as $A$ by connecting them to the
       generalized (correlation) functions (such as $G$) to which they give rise.
       Perhaps Monte Carlo calculations could make a contribution to
       that effort.

       In summary, in our all-too-brief discussion of the continuum limit,
       we have observed: (i) the atypical behavior of the
       limit in which $N'\ra\infty$, namely, in the case of generalized Poisson
       distributions, how, in hyper-spherical coordinates, the
       appearance of $N'$ as a large power of the overall field
       radius is absent, thereby rendering perturbation series
       termwise finite in contrast to a comparable calculation for
       Gaussian-like integrals; and (ii) the different role of
       internal parameters (e.g., matrices $A$ and $B$) in the
       ground-state distribution for Poisson- and Gaussian-like
       distributions, and how they effect the form taken by the
       continuum limit of correlation functions.

\section{Conclusions}
In discussing lattice quantum field theory models, it is
conventional -- and, of course, very natural -- to start with a
choice for the lattice action. A lattice action for a model such as
$\varphi^4_n$, $n\ge5$, however, necessitates that some difficult
choices must be made for the appropriate counterterms. Appropriate
in this sense means counterterms that have the virtue that the
resultant model and its continuum limit will faithfully describe the
original model chosen, as well as contain the original motivating
classical model in the limit that $\hbar\ra0$. This is a lot to ask,
and it is not surprising that the auxiliary potential (i.e.,
counterterm) generally chosen for this purpose does not perform as
desired.

In this paper we have taken a different approach. First of all,
recognizing that counterterms suggested by regularized, renormalized
perturbation theory are inappropriate for nonrenormalizable models
(see Sec.~2), we have felt compelled to direct our initial focus
away from the lattice action and place it instead on the sharp-time,
ground-state distribution function $\Psi(\p)^2$. To avoid
triviality, this distribution must be non-Gaussian, and to avoid the
basin of attraction that leads to a Gaussian distribution (and its
associated triviality), we have chosen to emphasize the possible
relevance of the only other class of suitable infinitely divisible
distributions, namely, the generalized Poisson distributions. Such
distributions avoid manifest triviality by being non-Gaussian, and
they are robust to survival -- as are the Gaussian distributions --
under general limiting conditions.

We found that ground-state distributions in general, and Poisson
distributions in particular, were faithful replacements for the
lattice action which implicitly -- even though somewhat indirectly
-- was determined by the ground-state distribution. Moreover, and
this is the important point, the class of lattice actions associated
with generalized Poisson distributions are automatically assured to
lead to manifest {\it non}triviality by ensuring a non-Gaussian
continuum limit.

Although we could be fairly specific about which Poisson
distribution we favored, it was nevertheless extremely difficulty to
see just which lattice action was implicitly connected with that
distribution. Thus we felt necessary to abandon any focus on
individual models, and rather focus on {\it classes} of models
hoping that such a wider net might capture one or another model of
possible interest.

In the putative continuum analysis of Sec.~3.4, it was emphasized
that suitable matrix elements of the canonical momenta, as well as
those of the Hamiltonian, were uniquely specified by our choice of
the characteristic function associated with the ground-state
distribution. It is interesting to further note that if it were
possible to actually determine the associated canonical coherent
states defined with the ground state itself serving as the fiducial
vector, then additional matrix elements of interest could be
computed. In particular, incorporating and extending the notation
for $|h\>$ of Sec.~3.4, we let
  \b |h,g\>\equiv e^{i\s\tint[h(x)\s\hph(x)-g(x)\s\hp(x)]\,d^{s}\!x}\,|\s0\> \e
for which it follows \cite{wcp} that
  \b \<h,g|\hph(x)|h,g\>=g(x)\;,\hskip1cm \<h,g|\hp(x)|h,g\>=h(x)\;, \e
and more significantly (assuming a $\varphi^4_n$ model under
consideration) that
  \b \lim_{\hbar\ra0}\,\<h,g|\s\H\s|h,g\>=\tint\{\s\half h(x)^2+\half[\nabla g(x)]^2+\half m^2\s g(x)^2
  +\lambda\s g(x)^4\s\}\,d^{s}\!x\;, \e
implying the close connection of the diagonal coherent state matrix
elements of the quantum Hamiltonian with the classical Hamiltonian
itself \cite {wcp}. All in all, these conditions place strong limits
on the ground-state wave function, even if they are highly implicit.

\section*{Acknowledgements}It is a pleasure to acknowledge the
hospitality of the Complex Systems and Soft Materials Research Group
of the Department of Physics, The Norwegian University of Science
and Technology, in Trondheim, as well as the support provided
through the 2006 Lars Onsager Professorship for an extended visit
during which time much of this paper was prepared. Thanks are also
extended to my principal host during this time period, Bo-Sture
Skagerstam.

\section*{Appendix: ~Fundamental Restrictions on the \\Choice of the
Weight Function $\rho(\p,a)$} In this Appendix we address the
question of acceptable values for $\g$ and acceptable choices for
the set $\{\B_k\}$ that are both part of the canonical form for
$\rho(\p,a)$ we have adopted in Sec.~4. Therefore, we again consider
the integral given by
 \b R(a)\int\s\frac{\p^{2p}_t\,e^{-U(\p,a)}}{\Pi'_k[\s\Sigma'_l\s\B_{k-l}\s\p_l^2\s]^\g}
 \,\Pi'_k\s d\p_k  \e
 which should {\it diverge} when $p=0$ and {\it converge} whenever $p\ge1$.
 We assume that $U(\p,a)$ controls large $\{\p_k\}$ behavior for all
 $p$, and so the question of divergence or convergence is related to
 the
 small $\{\p_k\}$ behavior where $U(\p,a)\simeq0$, and where $U(\p,a)$ is not a
 contributing factor. Indeed, for the general question of divergence
 or convergence the specific form of $U(\p,a)$ is not of particular
 relevance and it may effectively be replaced by
   \b U(\p,a)=\Sigma'_k\p_k^2\s a^s\;,  \e
leading to the expression of interest
  \b \hskip.5cm I_p\equiv R(a)\int\s\frac{\p^{2p}_t\,e^{-\Sigma'_k\p_k^2\s a^s}}
  {\Pi'_k[\s\Sigma'_l\s\B_{k-l}\s\p_l^2\s]^\g}
 \,\Pi'_k\s d\p_k\;,\hskip2cm p=0,1,2,\ldots\;.  \e

It is convenient at this point to again introduce hyper-spherical
coordinates for which
 $\p_k\equiv \k\s\eta_k$, $\k\ge0$, $-1\le\eta_k\le1$,
 where $\Sigma'_k\s\eta_k^2\equiv1$, and $\Sigma'_k\s\p^2_k=\k^2$.
In terms of these coordinates, the integral of interest reads
 \b I_p=2R(a)\int\frac{\k^{2p}\s\eta_t^{2p}\,e^{-\k^2\s a^s}\s\k^{N'-1}\,d\k\s
 \delta(1-\Sigma'_k\eta_k^2)\s\Pi'_k\s d\eta_k}{\k^{2\g
 N'}\,\Pi'_k[\s\Sigma'_l\B_{k-l}\s\eta^2_l\s]^\g}\;.\e
For $p=0$ we attribute divergence of this integral to a divergence
at $\k=0$. Consequently, we need $2\g N'\ge N'$, i.e., $\g\ge 1/2$.
For $p\ge1$, convergence at $\k=0$ requires that $2\g N'<N'+2$,
i.e., $\g<1/2+1/N'$. Thus we are led to the bounds \b
\frac{1}{2}\le\g< \frac{1}{2}+\frac{1}{N'}\;.
 \e Since the limit
$N'\ra\infty$ is eventually to be taken, we already satisfy this
situation by adopting $\g=1/2$, as noted earlier. However, we are
not quite finished because convergence for $p\ge1$ also requires
convergence of the integrals over the $\{\eta_k\}$, and as we now
shall see, this convergence requires some restriction on the set of
coefficients $\{\B_k\}$.

The first remark regarding such integrals is to observe that with
$\g=1/2$ it is not possible that $\B_k\propto\delta_{k0}$, for in
that case integrals over each $\eta_k$, save when $k=t$, lead to a
divergence. Consequently, $\B_k$ must be nonvanishing for additional
sites. To study how many additional sites are sufficient, it is
convenient to express our basic integral in yet another form, namely
  \b && I_p=R(a)\int\s\frac{\p^{2p}_t\,e^{-\Sigma'_k\p_k^2\s
 a^s}}{\Pi'_k[\s\Sigma'_l\s\B_{k-l}\s\p_l^2\s]^{1/2}}
 \,\Pi'_k\s d\p_k\\
 &&\hskip.46cm=R(a)\, a^{s N'/2}\s\pi^{-N'/2}\int\p_t^{2p}\s e^{-\Sigma'_k\p_k^2\s
 a^s}\s\Pi'_k\s d\p_k\\
 &&\hskip2cm\times\int
 [\s\Pi'_k\l^{(-1/2)}_k\s]\,e^{-\Sigma'_k\l_k[\s\Sigma'_l\B_{k-l}\p_l^2\s]\s
 a^s}\,\Pi'_k\s d\l_k\\
 &&\hskip.46cm=R(a)\, a^{s N'/2}\s\pi^{-N'/2}\int\!\int\p_t^{2p}\s[\s\Pi'_k\l^{(-1/2)}_k\s]\,
 e^{-\Sigma'_k(1+\Sigma'_l\B_{k-l}\l_l\s)\p_k^2\s
 a^s}\,\Pi'_k\s d\l_k\,\Pi'_k\s d\p_k\;,\e
 where the integral for each $\l_k$ variable runs from $0$ to
 $\infty$. Exchanging the order of integration leads to
  \b I_p\equiv K\int\frac{[\s\Pi'_k\l_k^{(-1/2)}\s]\,\Pi'_k\s
  d\l_k}
  {[1+\Sigma'_l\B_{t-l}\s\l_l]^p\,[\Pi'_k(1+\Sigma'_l\B_{k-l}\s\l_l)^{1/2}]}\;,\e
where
  \b K\equiv 2^{-p}\s(2p-1)!!\s R(a)\,
  a^{-sp}\;.\e
Observe that when the basic integral $I_p$ is expressed in the
$\{\l_k\}$ variables, divergence or convergence concerns the
behavior of the integrand for {\it large} $\{\l_k\}$.

Initially, let us again show, when  $p=0$, that $I=\infty$  -- this
time in terms of the $\{\l_k\}$ integration variables. This result
follows from the lower bound given by

 \b I_0\ge K\int\frac{[\s\Pi'_k\l_k^{(-1/2)}\s]\,\Pi'_k\s
  d\l_k}{[\Pi'_k(1+\S_l\l_l]^{1/2}}\;,
  \e
which holds because $\B_k\le1$. If we introduce coordinates of the
form $\l_k\equiv\zeta\eta^2_k$, $\zeta\ge0$, $-1\le\eta_k\le1$, and
$\S_k\eta_k^2=1$, then it follows that

 \b I_0\ge 2^{N'}\s K\int\frac{\zeta^{(N'/2-1)}\,d\zeta}
  {[1+\zeta]^{N'/2}}\,\int\,\delta(1-\S_k\eta^2_k)\,\Pi'_k\s
  d\eta_k\;.
  \e
 The first integral diverges, while the
  remaining factors are positive. Thus we have again shown that
$I_0=\infty$, as expected.

For $p\ge1$, we require that $I_p<\infty$. First we observe by
construction that $I_q\le I_p$ whenever $q\ge p$. Therefore, to show
that $I_p<\infty$ for $p\ge1$, it suffices to show that
$I_1<\infty$. Thus we focus attention on

\b I_1\equiv K\int\frac{[\s\Pi'_k\l_k^{(-1/2)}\s]\,\Pi'_k\s
  d\l_k}
  {[1+\S_l\B_{t-l}\s\l_l]\,[\Pi'_k(1+\S_l\B_{k-l}\s\l_l)^{1/2}]}
\e

If this integral were to diverge it must diverge for large
$\lambda$. We can reach large $\lambda$ in many different
directions. In particular, we can imagine that $P$, $1\le P\le N'$,
of the $\lambda$ variables are all approaching infinity in a certain
direction. As an example,  consider $P=3$ and $\lambda_1=.5\s\zeta$,
$\lambda_2=.2\s\zeta$, $\lambda_3=.1\s\zeta$, as $\zeta\ra\infty$,
while all other $\lambda$ variables have finite values. The
``direction'' of approach to infinity in this case refers to the
relative size of the growing $\lambda$ terms, i.e., the direction as
determined by $\eta_1^2=.5$, $\eta_2^2=.2$, and $\eta_3^2=.1$.
Staying with this example for a moment, the three large $\lambda$
values will arise in several terms in the denominator because of the
possible many-fingered nature of the $\beta$ terms. We divide the
set of distinct space-like $k$ values into the set $P$ which
contains one or more of the $P$ variables approaching infinity and
the set $Q$ containing the remaining $N'-P$ terms for which the
variables $\l$ do not approach infinity. In dividing the appropriate
sets below, we have added an asterisk to the set of $\l_k$ values
that are approaching infinity to serve as a reminder of that fact.
Thus we are led to the expression \b
&&I_1=K\int\int\frac{[\s\Pi'_{k\in P}\l_k^{*\s(-1/2)}\s]}
{[1+\S_{l\in P}\B_{t-l}\s\l_l^*+ \S_{l\in Q}\B_{t-l}\s\l_l]}\\
&&\hskip2cm\times\frac{[\s\Pi'_{k\in
Q}\l_k^{\s(-1/2)}\s]\,\Pi'_{k\in P}\s
  d\l_k^*\;\Pi'_{k\in Q}\s  d\l_k}
{\,[\Pi'_k(1+\S_{l\in P} \B_{k-l}\s\l_l^*+\S_{l\in Q}
\B_{k-l}\s\l_l)^{1/2}]}\;. \e

In the numerator of this expression there are $P$ variables
approaching infinity at the same time, where $1\le P\le N'$, and we
need to consider all possibilities. For the denominator, there are
several cases to consider. We assume that the set $\{\B_k\}$ has
$1+S_0$ nonvanishing terms and that they are connected to the ``home
coordinate'', $k=0$. In particular, let us focus on the choice $\kk$
(consisting of the point $k$ and its $2s$ nearest neighbors in
 spatial directions) for which $S_0=2s$. As a consequence, there are large $\l$ variables
in $P+S\,(+1)$ factors in the denominator; here the term $S$, which
may vary from case to case, arises from the many-fingered nature of
$\{\B_k\}$, and the term $(+1)$ adds $1$ only if a large $\l$ factor
appears in the denominator factor associated with $p=1$. Convergence
of $I_1$ is ensured if for all $P$, $1\le P\le N'$, the choice of
$\{\B_k\}$ ensures that $S\,(+1)\ge1$ in all possible combinations
of the large $\l$ values. A few specific examples may help clarify
how the choice of $\kk$ fulfills all the necessary requirements.

For example, if $P=1$ then $S\,(+1)=S_0\,(+1)\ge1$. If $P=N'-1$
includes all the lattice points except one, then $S\,(+1)=1$; if
$P=N'$, and thus includes all lattice points, then $S\,(+1)=1$. If
$P=L$, all of which lie in a single coordinate direction (with
period $L$), then $S\,(+1)=2(s-1)L\,(+1)\ge1$. Incidentally, this
example shows that fingering in just {\it one} spatial direction is
{\it not} sufficient -- and, by analogy, fingering in $(s-1)$
spatial directions is also insufficient. It is necessary to finger
in all $s$ different spatial directions.

A picture may also help; see Fig.~1. Although not directly relevant
to nonrenormalizable fields, we give a two-dimensional example
($s=2$) for which the points with large $\l$ values are marked with
solid circles
$\begin{picture}(5,5)(-2,-3)\circle*{6}\end{picture}\s$, the
additional points in which those large values appear in denominator
factors are marked with an open circle
$\begin{picture}(5,5)(-2,-3)\circle{6}\end{picture}\s$,
  and the point $k=t$ is marked with an $\times$. In the example of this picture, $P=2$, $S_0=4$, $S=6$,
  and since the $p=1$ term for this example does not contribute, the $(+1)$ term is absent.

\begin{picture}(300,240)(40,10)
\put(165,165){\line(1,1){10}} \put(165,175){\line(1,-1){10}}
\put(210,50){\circle{9}} \put(210,90){\circle*{9}}
\put(210,130){\circle{9}} \put(250,50){\circle{9}}
\put(250,90){\circle*{9}} \put(250,130){\circle{9}}
\put(170,90){\circle{9}} \put(290,90){\circle{9}}
\multiput(130,10)(0,40){6}{\line(1,0){200}}
\multiput(130,10)(40,0){6}{\line(0,1){200}}
\end{picture}
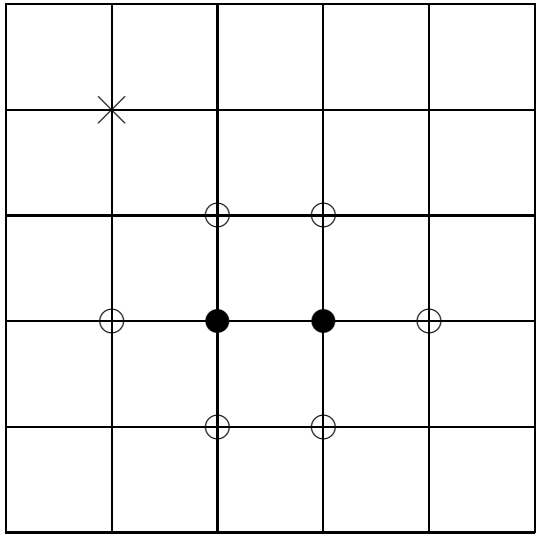
\vskip.8cm \indent {Fig.~1: A two-dimensional example with two large
$\l$ values (dark circles)

which also have a presence in six additional locations (open
circles). The

place marked by $\times$ is where the additional moment is located.}

\end{document}